\documentclass[]{spie}  
\usepackage[]{graphicx}
\usepackage{amsmath,amssymb}
\usepackage{float}
\usepackage[bf]{caption}
\title{Xampling in Ultrasound Imaging} 

\author{Noam Wagner\supit{a}, Yonina C. Eldar\supit{a,b}, Arie Feuer\supit{a}, Gilad Danin\supit{a} and Zvi Friedman\supit{c}
\skiplinehalf
\supit{a}Technion-Israel Institute of Technology, Technion City, Haifa, Israel; \\
\supit{b}Visiting Professor, Electrical Engineering Department, Stanford, CA; \\
\supit{c}GE Health-care, Technion City, Haifa, Israel
}

\authorinfo{Further author information: (Send correspondence to Noam Wagner)\\Noam Wagner: E-mail: noamwa@tx.technion.ac.il, Telephone: +972(0)73 725 2653
\\Yonina C. Eldar: E-mail: yonina@ee.technion.ac.il, Telephone: +972(4) 829 3256
\\Arie Feuer: E-mail: feuer@ee.technion.ac.il, Telephone: +972(4) 829 4648
\\Gilad Danin: E-mail: daning@tx.technion.ac.il, Telephone: +972(4) 829 4706 
\\Zvi Friedman: E-mail: zvi.friedman@med.ge.com}

\begin{document} 
  \maketitle 

\begin{abstract}
\noindent Recent developments of new medical treatment techniques put challenging demands on ultrasound imaging systems in terms of both image quality and raw data size.  Traditional sampling methods result in very large amounts of data, thus, increasing demands on processing hardware and limiting the flexibility in the post-processing stages.    

\noindent In this paper, we apply Compressed Sensing (CS) techniques to analog ultrasound signals, following the recently developed Xampling framework.  The result is a system with significantly reduced sampling rates which, in turn, means significantly reduced data size while maintaining the quality of the resulting images.  
\end{abstract}


\keywords{Array Processing, Beamforming, Compressed Sensing, Dynamic Focus,  Finite Rate of Innovation (FRI), Matrix Pencil, Ultrasound,  Xampling}

\section{INTRODUCTION}
\label{sec:intro}  
\noindent Modern ultrasound systems utilize an array of transducer elements in a process known as beamforming~\cite{Szabo01, Jensen02}.  An imaging cycle begins when modulated acoustic pulses are transmitted from some or all array elements.  Specific delays are applied to the transmitting elements, such that the interfering waves form a narrow beam, along which most energy propagates.  As the energy propagates, the beam gets narrower, until reaching the focal point, after which it expands.  Determining the position of the focal point is achieved by applying appropriate delays to the transmitting elements.  

\noindent As the focused energy propagates along the tissue, echoes are scattered and reflected by density and propagation velocity perturbations \cite{Jensen01}.  The array elements detect the reflected energy.  A second beamforming process is then performed, aimed at localizing the reflecting elements, while at the same time improving signal to noise ratio (SNR).  This beamforming process is performed by summing delayed samples of the received data.  For that purpose, high rate A/D conversion must be first carried out at each of the receiving channels.  Actually, optimizing the image resolution requires that each receiving channel will be sampled at 3-5 times the center frequency of the modulated pulse \cite{Szabo01}.  More explicitly, let us consider a B-Mode scan, in which the central frequency of the transducer may vary in between  $2-15_{MHz}$, depending on the use\cite{Jensen02}.  Advanced composite materials, often used in the transducer, can attain a relative bandwidth in excess of 100\%.  Therefore, if we assume a nominal center frequency of $5_{MHz}$, we end up with a baseband bandwidth of approximately $10_{MHz}$.  Confined to the classic Nyquist-Shannon sampling theorem~\cite{Shannon01}, where the only prior on the signal is that it is bandlimited, this implies that standard ultrasound devices must sample the analog signal received in each active element at a typical rate of at least $20_{MHz}$.    

\noindent Recent works~\cite{Vetterli01, Mishali01, Mishali03, Mishali04, Matusiak01} show that, by exploiting other priors regarding the signal structure, it is possible to design more efficient sampling schemes, which actually break through the Nyquist barrier.  Decreasing the sampling rate is of vast interest, as it may obviously be manifested in reduction of machinery size and power consumption.   

\noindent In their recent work, Tur, Eldar and Friedman~\cite{Tur01} first proposed to implement these ideas in ultrasound imaging.  They suggested to regard the signal received in each of the array transducer elements as having finite rate of innovation (FRI)\cite{Vetterli01}.  More specifically, they assume each such signal comprises at most $L$ replicas of a known-shape pulse, all received within the time interval $\left[0,\tau\right)$.   Delay and gain parameters are associated with each replica, such that the entire problem may be  characterized by $2L$ degrees of freedom. They then developed a new sub-Nyquist sampling technique that exploits this structure, in order to reduce the sampling rate way beyond that which is used in standard ultrasound devices, dictated by the classic Nyquist-Shannon sampling theorem.

\noindent Tur, Eldar and Friedman's work follows the spirit of analog compressed sensing, also referred to as Xampling~\cite{Mishali01}.  The latter is an emerging framework, which combines classic methods from sampling theory together with recent developments in compressed sensing, aimed at sampling analog signals far below the Nyquist rate.  Throughout this paper, we will use the term Xampling whenever referring to the sub-Nyquist sampling schemes purposed in Ref.~\citenum{Tur01}, and in later work by Gedalyahu, Tur and Eldar~\cite{Gedalyahu01}.  It should be noted, though, that both these schemes are special cases of Xampling, in which the signal's FRI property is exploited in order to achieve the goal of sub-Nyquist sampling.           

\noindent The Xampling scheme suggested by Tur, Eldar and Friedman~\cite{Tur01} comprises the following steps: first, the received signal is filtered using the compactly supported Sum of Sincs kernel.  The filtered signal is then sampled at nearly the rate of innovation, namely the number of unknown parameters $2L$, which is much smaller than the Nyquist rate of the pulse.  The extracted samples are used for computing a finite set of Fourier coefficients, which correspond to the $\tau$-periodic extension of the received signal.  Having obtained the finite set of Fourier coefficients, spectral analysis techniques, similar to these presented by Vetterli et al.~\cite{Vetterli01,Blu01}, are utilized, in order to estimate the set of unknown delays and amplitudes characterizing the reflected pulses. 

\noindent Whereas Tur, Eldar and Friedman\cite{Tur01} adopt a filtering and sampling approach as a preliminary step for obtaining the required set of Fourier coefficients, Gedalyahu, Tur and Eldar's\cite{Gedalyahu01} later work suggests a second approach for obtaining the same coefficients: the single filtering and sampling channel is replaced by a bank of modulators and integrators.  Both approaches are aimed at minimal rate sampling of a {\bf{single}} received channel.  

\noindent Referring to an array of transducer elements utilized in the ultrasound imaging device, by processing the signal received in each channel separately (using either approach), one may obtain a corresponding set of delays and amplitudes from low-rate samples.  The sets obtained from all receiving elements may then be combined (via some geometric interpretation), in order to estimate the two-dimensional coordinates of the reflecting elements.  Note, however, that such schemes cannot achieve the SNR improvement which is an integral part of standard beamforming techniques; this is because the correlation between signals received in different channels is not exploited throughout the process of extracting the parameters (pulse amplitudes and delays) from each signal.  More explicitly, the Xampling schemes proposed in Refs.~\citenum{Tur01} and~\citenum{Gedalyahu01} both aim at accurately detecting strong, localized pulses, related with macroscopic perturbations.  However, the actual signals received by the array elements also contain components which, in the context of our work, may be regarded as noise.  Actually, part of these noisy components arises from constructive and destructive interference of acoustic waves reflected from dense, subwavelength scatterers in the tissue (these are typically manifested as granular texture in the ultrasound image, called ‘‘speckle’’, after a similar effect in laser optics\cite{Szabo01}).  Apparently, these noisy components induce erroneous results when Xampling the received signals.  If we wish to obtain meaningful results by either Xampling scheme, while maintaining a rather low oversampling factor, the overall SNR improvement is indeed a crucial step.  

\noindent This paper is aimed at generalizing the schemes of Refs.~\citenum{Gedalyahu01} and~\citenum{Tur01} to multiple antenna arrays.  Our goal is to obtain a two-dimensional, focused ultrasound image, corresponding to strong perturbations in the scanned plane, while reducing the sampling rate in each active element, by a factor of 10-15 times relative to the rate used in standard ultrasound devices.  Furthermore, we aim at achieving this goal in the presence of noise in the received signals.  In such case, straightforward implementation of either schemes of Refs.~\citenum{Tur01} or~\citenum{Gedalyahu01} on each channel independently, would require hard thresholding (which in turn cancels/attenuates desired pulses) and/or increasing the oversampling factor, such that the final sampling rate grows towards the Nyquist rate.  

\noindent Our Xampling scheme's most expensive computational component regards the extraction of the pulses' delays and amplitudes from the set of low rate samples.  Referring to standard ultrasound devices, this component substitutes both  the Hilbert transform, which is applied throughout the process of envelope detection\cite{Szabo01}, and the expensive beamforming computations, namely: summing samples obtained at Nyquist rate from all active elements.  In addition, referring to straightforward implementation of either systems of Refs.~\citenum{Tur01} or~\citenum{Gedalyahu01} on each channel separately, our scheme extracts the pulses' parameters once per image line, rather than once per active element participating in the image line generation (which equivalently means tens of times per image line, depending on the number of active elements used for beamforming).   

\noindent The Xampling system we propose may be summarized as follows:  Let us assume that we could somehow generate the beamformed signal, corresponding to a single image line, in the analog domain.  Assuming that such a signal maintains the FRI property of the received signals from which it was constructed, we may Xample it using the scheme suggested by either Ref.~\citenum{Tur01}  or~\citenum{Gedalyahu01}, yielding the delays and amplitudes of pulses along the final beamformed image line.  By mathematically formulating the process of Xampling the beamformed signal using the scheme suggested in Ref.~\citenum{Gedalyahu01}, and then applying several algebraic manipulations to this formulation, we end up with a new set of generalized modulation kernels.  The latter may now be applied directly to the analog signals which are received in each of the active elements, thus bypassing the impractical step of actually generating the beamformed signal in the analog domain.
      
\noindent Due to space limitation, the following paper is not aimed at presenting a rigorous review of our results.  Instead, we present an outline of our approach, and preliminary results obtained using actual ultrasound data.  The paper is organized as follows:  Section \ref{sec:02} briefly outlines principles of standard ultrasound imaging, namely the process of beamforming in both Polar and Linear scan methods, applying dynamic receive (Rx) focus.  Section \ref{sec:03} reviews the one-dimensional Xampling scheme suggested by Refs.~\citenum{Tur01} and~\citenum{Gedalyahu01}.  In section \ref{sec:04} we present our system, which combines the concepts of beamforming with those of one-dimensional Xampling.   Section \ref{sec:05} provides results obtained by applying our Xampling scheme upon actual ultrasound data.  In this section we also analyze the reduction in the necessary amount of samples, and the way this affects the overall computational cost.  Finally, conclusions are drawn in Section \ref{sec:06}. 
\section{Dynamic Focusing in Polar and Linear Scan} 
\label{sec:02}  
\noindent This section is aimed at outlining the method by which an ultrasound image is generated using a linear array of transducer elements.  Our discussion refers mainly to B-mode scan, in which the array simultaneously scans a plane through the body, resulting in a two-dimensional image, which may be viewed on screen.  We are specifically interested in the process carried out by the electronic beamformer, in which multiple signals, received from a set of receivers, are focused into a single trace, known as the beamformed signal.  The latter is envelope detected, forming a single image line. The analysis reviewed throughout this section is based mainly on Ref.~\citenum{Jensen01}.

\noindent An ultrasound image consists of roughly 100 lines.  Each line is generated throughout a single transmit-receive cycle.  In such a cycle, a set of active elements first transmits acoustic pulses, modulated to central frequency of $2$ to $15_{MHz}$, depending on the use.  An appropriate delay is applied to the pulse transmitted from each transducer element, aimed at obtaining constructive interference at a specific point in the plane, referred to as the focal point.  The interfering acoustic waves form an acoustic pulse, which propagates along a narrow beam (containing most of the transmitted energy).  The beam gets narrower, until reaching the focal point, after which it expands.  

\noindent The velocity at which the pulse propagates will be denoted by $c$, and varies between $1446_{m/sec}$ (fat) to $1566_{m/sec}$ (spleen).  An average value of $1540_{m/sec}$ is assumed by scanners for processing purposes.   As the transmitted pulse propagates inside the tissue, it encounters density and propagation-velocity perturbations.  These cause scattered and reflected echoes, which are detected by the array elements.  Applying the acoustic reciprocity theorem~\cite{Kinsler01}, a second beamforming process is now carried out, in which the received signals are combined into a single trace, which in a sense, visualizes structures in the tissue, along the transmitted beam.  Modern ultrasound devices typically handle 64 to 192 transducer elements in the beamforming process.  We emphasize, that combining the received signals is performed in the digital domain, implying that modern ultrasound devices must first sample the signal received in each of the active elements at the Nyquist rate (typically $20_{MHz}$).  Refs.~\citenum{Tur01} and~\citenum{Gedalyahu01}, introduced an approach, which allowed reconstruction of the analog signal detected by an individual transducer element from a very low number of samples.  In contrast, our generalized scheme is aimed at obtaining the low rate samples from all active elements, in a manner which will enable to directly reconstruct the  beamformed signal.  This is further discussed in Section~\ref{sec:04}. 

\noindent We would now like to formulate the manner in which the beamformer combines the signals received in all active elements into a single trace, known as the beamformed signal, and better understand the significance of the latter.  This formulation applies to conventional ultrasound imaging, and will be necessary when we translate our  theoretical scheme of Xampling the beamformed signal, into an applicable scheme where the signals received in the active transducer elements are sampled directly.  

\begin{figure}
   \begin{center}
   \begin{tabular}{c}
   \includegraphics[height=7cm]{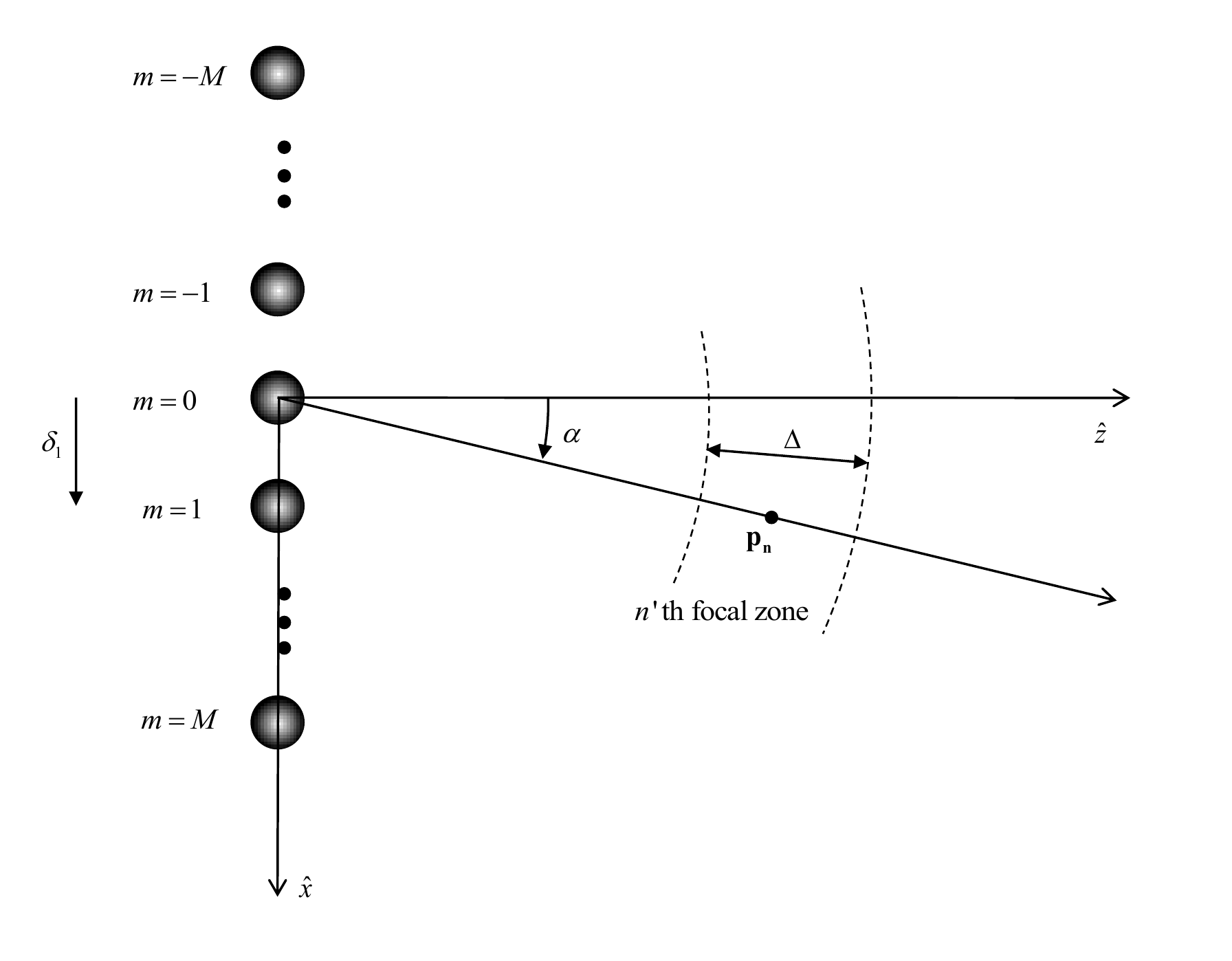}
   \end{tabular}
   \end{center}
   \caption[01]
   {\label{Fig:01} 
$2M+1$ elements aligned along the ${\bf{\hat{x}}}$ axis.  The ray along which the pulse propagates forms an angle $\alpha$ with ${\bf{\hat{z}}}$ axis.  We analyze a reflection emerging from the point ${\bf{p_n}}$, which is associated with the $n$th focal zone.}    
   \end{figure}

\noindent Referring to Figure~\ref{Fig:01}, we examine the two-dimensional plane XZ in which $2M+1$ elements are aligned along the ${\bf{\hat{x}}}$ axis (the center of the array coincides with the origin).  We analyze one cycle, in which a single 
image line is constructed.  The image line corresponds to a beam located within the XZ plane, emerging from the array center.  We denote by $\alpha$, the angle between the beam and ${\bf{\hat{z}}}$ axis (normal to the array).  

\noindent The cycle begins, when each active element transmits a single modulated pulse, such that the interference pattern may be observed as a concentrated pulse of energy, propagating along the beam.  We regard the pulse as if it was transmitted from the array center, at a known time, which we shall define as $t=0$.  Knowing the speed in which the pulse propagates (denoted by $c$, and assumed $1540_{m/sec}$), we may now estimate the distance which it traveled along the beam by the time instance $t_n$, denoted by $r(t_n)$, and thereby its two-dimensional position, ${\bf{p_n}}$:  
\begin{equation}\label{E:01}
{\bf{p}_n}=\left[\begin{array}{ll} \mbox{cos}\alpha &  \mbox{sin}\alpha \end{array}\right]^T ~r\left(t_n\right)=\left[\begin{array}{ll} \mbox{cos}\alpha &  \mbox{sin}\alpha \end{array}\right]^Tct_n.
\end{equation}

\noindent Assuming that an echo was reflected due to some perturbation located at ${\bf{p_n}}$, we may easily estimate the time in which it will arrive back at the origin, $\tau_{0}\left(t_n\right)$:
\begin{equation}\label{E:02}
\tau_{0}\left(t_n\right)=t_n+\frac{1}{c}||{\bf{p}_n}||=2t_n. 
\end{equation}
 
\noindent Since the distance from ${\bf{p_n}}$ to each of the active elements varies, each element will detect the reflected pulse at a different time instance.  Namely, the time at which the pulse arrives at the $m$th receiver, positioned at ${\bf{x_m}}$, is given by:
\begin{equation}\label{E:03}
\tau_{m}\left(t_n\right) =t_n+\frac{1}{c}||{\bf{p}_n}-{\bf{x}_m}||=t_n+\frac{1}{c}\sqrt{\left(ct_n\mbox{sin}\alpha-\delta_m\right)^2+\left(ct_n\mbox{cos}\alpha\right)^2},
\end{equation}    
where $\delta_m$ denotes the $x$ coordinate of the receiving element.

\noindent Let us denote by $\varphi_m(t)$, the analog signal detected by the active element indexed $m$.  Beamforming is now achieved by shifting each of the received signals $\varphi_m(t)$, in order to compensate for the time difference $\tau_m(t_n)-\tau_0(t_n)$, and then summing the shifted versions. The acoustic reciprocity theorem~\cite{Kinsler01} implies that when we sum the shifted signals, constructive interference will occur at $t=\tau_0(t_n)=2t_n$, providing that an echo was indeed reflected from ${\bf{p_n}}$ at time $t_n$.  Denoting by $\Phi(t;\alpha)$ the sum of the delayed signals, we are thus interested in the value which $\Phi(t;\alpha)$ obtains at $\tau_0(t_n)=2t_n$.   

\noindent Shifting the signal $\varphi_m(t)$ so that the difference $\tau_m(t_n)-\tau_0(t_n)$ is compensated, is obtained by applying the (possibly negative) delay
\begin{equation}\label{E:04}
\theta_{m}\left(t_n;\alpha\right) =\tau_{0}(t_n)-\tau_{m}(t_n)=t_n-\sqrt{t_n^2+\left(\delta_m/c\right)^2-2t_n\left(\delta_m/c\right)\mbox{sin}\alpha}
\end{equation} 
to the signal received in the $m$th element.  

\noindent Summarizing the above, we have: 
\begin{align}\label{E:05}
\Phi(2t_n;\alpha)&=\Phi(t;\alpha)|_{t=2t_n}=\sum_{m=-M}^M{{\varphi}_m\left(t-\theta_{m}\left(t_n;\alpha\right)\right)}|_{t=2t_n}\notag\\
&=\sum_{m=-M}^M{{\varphi}_m\left(t_n+\sqrt{t_n^2+\left(\delta_m/c\right)^2-2t_n\left(\delta_m/c\right)\mbox{sin}\alpha}\right)},
\end{align}          
which corresponds to the intensity of a reflection originating at time $t_n$, from the coordinate ${\bf{p_n}}$.  For purposes of convenience, we will finally substitute $2t_n\rightarrow t$, obtaining an expression for the beamformed signal:
\begin{equation}\label{E:06}
\Phi(t;\alpha)=\sum_{m=-M}^M{{\varphi}_m\left(\frac{1}{2}\left(t+\sqrt{t^2+4\left(\delta_m/c\right)\left(\left(\delta_m/c\right)-t\mbox{sin}\alpha\right)}\right)\right)}.       
\end{equation}
\noindent Referring to~(\ref{E:01}), our last substitution implies that $\Phi(t;\alpha)$ represents the intensity of a reflection originating at time $t/2$ from a point distanced $ct/2$ from the origin, along the transmitted beam.  The assumption is that the transmitted pulse indeed intersected this point at $t/2$, and was possibly scattered.  $\Phi(t;\alpha)$ is obviously the result of varying the receive focal point along time, and its construction is therefore referred to as  ``Dynamic Focusing".  

\noindent We note that, instead of obtaining the beamformed signal using dynamic focusing process, beamformers sometimes divide the image line into $N$ segments, called focal zones, such that an entire segment of $\Phi(t;\alpha)$, corresponding to the $n$th focal zone, is constructed using a single set of delays (one delay per element).  The set is obtained using~(\ref{E:04}), which is calculated for a single, representative  point within the focal zone.     

\noindent In this section, we have formulated the manner in which the beamformer combines the signals received in the transducer elements, $\left\{\varphi_m(t)\right\}_{m=-M}^M$ into a beamformed signal, $\Phi(t;\alpha)$, through the process of dynamic focusing.  The dynamically focused, beamformed signal, may now be used in order to generate a single image line.  The beamformer performs the computation formulated in~(\ref{E:06}) in the digital domain, using samples obtained from each of the transducer elements at the Nyquist rate.  Our goal is to retrieve a set of parameters from which the beamformed signal $\Phi(t;\alpha)$ formulated in~(\ref{E:06}) may be reconstructed, by sampling the received signals far below the Nyquist rate.  We achieve this by exploiting an approximately FRI structure characterizing $\Phi(t;\alpha)$, within the Xampling methods of Refs.~\citenum{Tur01} and~\citenum{Gedalyahu01}, which are outlined in the next section. 

\noindent Throughout the rest of this paper, we will limit ourselves to the case of linear scan, in which all beams are parallel to the ${\bf{\hat{z}}}$ axis.  This is achieved by setting $\alpha=0$ in~(\ref{E:06}).  The parameter $\delta_m$ now represents the distance between the $m$th receiver and the beam processed at the current cycle.  Equation~(\ref{E:06}) then becomes:
\begin{equation}\label{E:07}
\Phi(t;\alpha=0)=\sum_{m=-M}^M{{\varphi}_m\left(\frac{1}{2}\left(t+\sqrt{t^2+4\left(\delta_m/c\right)^2}\right)\right)}.       
\end{equation}

\section{Xampling the Signal obtained in a single Transducer Element} 
\label{sec:03}  
\noindent Up until this point, we have outlined the process in which a B-mode ultrasound image line is generated.  The cycle begins by transmitting a modulated pulse along a narrow beam.  The device then captures the intensity of echoes reflected along the beam by applying dynamically focused beamforming.    

\noindent Regarding the ultrasound signal detected by a single transducer element indexed $m$ (denoted by $\varphi_m\left(t\right)$), Tur, Eldar and Friedman~\cite{Tur01} assume that it consists of a set of known-shape pulses, which result from reflections of the transmitted pulse by strong, macroscopic perturbations in the tissue.  The signal can hence be approximated as an FRI of the form:
\begin{equation}\label{E:08}
\varphi_m(t)=\sum_{l=1}^L{a_{l,m} h(t-t_{l,m} )},
\end{equation}
\noindent where $h(t)$ is a known-shape pulse, and there exists some $\tau>0$ such that $h(t-t_{l,m})=0$, $\forall t\notin\left[0,\tau\right)$, $l=1...L$.      

\noindent The extremely short support of $h(t)$ implies that $\varphi_m(t)$ is of very wide band (the support of $h(t)$ is typically $600_{nsec}$, with $\tau$ being typically $200_{\mu sec}$).  Classic Nyquist-Shannon sampling theorem thus forces standard ultrasound devices to sample $\varphi_m\left(t\right)$ at a high rate (typically $20_{MHz}$).  Nevertheless, we may easily observe that $\varphi_m(t)$ actually has only $2L$ degrees of freedom ($L$ being the number of macroscopic scatterers along the path of the transmitted pulse).  The schemes introduced by Refs.~\citenum{Tur01} and~\citenum{Gedalyahu01} manage to exploit this property, enabling the reconstruction of $\varphi_m\left(t\right)$ from a much smaller number of samples (at least $2L$ per time interval $\tau$). 

\noindent In this section we outline the system suggested by Ref.~\citenum{Gedalyahu01} for Xampling the signal $\varphi_m(t)$.  This approach is the basis for our scheme, introduced in Section \ref{sec:04}.  We emphasize that both Refs.~\citenum{Gedalyahu01} and~\citenum{Tur01} treat Xampling of a one dimensional signal which is received in a single transducer element.  Moreover, they do not treat integration of samples, obtained from multiple elements, into a two-dimensional ultrasound image.  Our novelty concerns a method for obtaining parameters of the {\bf{beamformed signal}}, $\Phi(t;\alpha=0)$  (introduced in the previous section), which is directly related to the image line.  We obtain these parameters from low rate samples of the individual signals $\varphi_m(t)$.  By extracting the parameters of the beamformed signal we also cope with the noisy components, which induce erroneous results when attempting to reconstruct the individual signal $\varphi_m(t)$, from its corresponding low rate samples.  

\begin{figure}
   \begin{center}
   \begin{tabular}{c}
   \includegraphics[height=7cm]{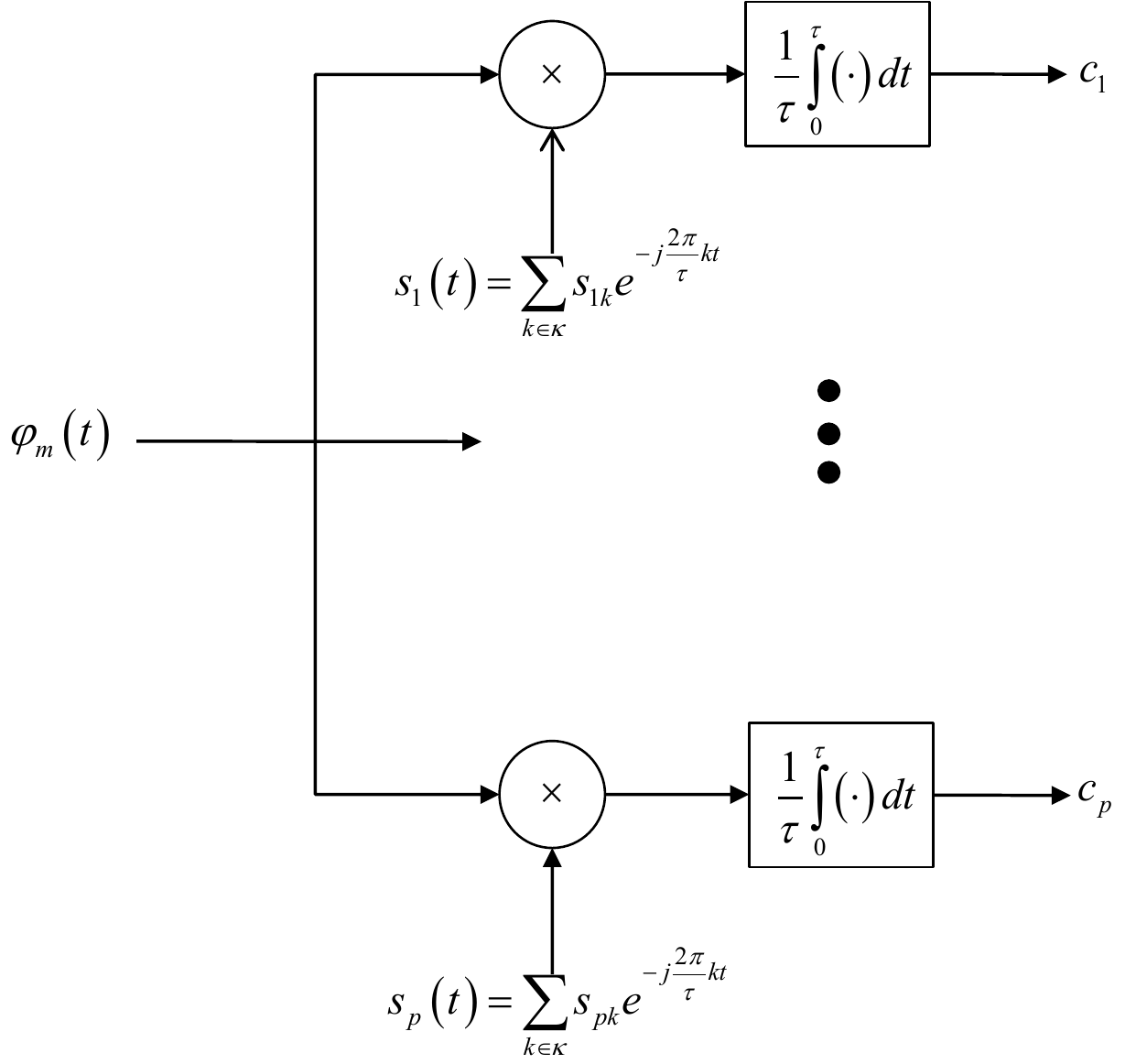}
   \end{tabular}
   \end{center}
   \caption[02]
   {\label{Fig:02} Gedalyahu, Tur and Eldar's~\cite{Gedalyahu01} multichannel Xampling scheme.  The resulting samples $\left\{c_{q}\right\}_{q=1}^p$ constitute a mixture of Fourier coefficients corresponding to $\varphi_m(t)$'s $\tau$-periodic extension.  The coefficients' indices belong to the set $\kappa$.  Figure reprinted from Ref.~\citenum{Gedalyahu01}.}    
   \end{figure}

\noindent Let us denote by $H(\omega)$ the CTFT of the known-shape pulse $h(t)$, and by $\phi_m[k]$ the $k$th Fourier coefficient of $\varphi_m(t)$'s $\tau$-periodic extension.  Further denote by $\kappa$, a set of $K$ consecutive indices for which $H(\omega=\frac{2\pi}{t}k)\neq0$, $\forall k\in\kappa$.  Ref.~\citenum{Tur01} shows that, as long as $K\geq2L$ and the unknown time delays are distinct, i.e. $t_{i,m}\neq t_{j,m}$, $\forall i\neq j$, one may accurately estimate $\varphi_m(t)$, from the set $\left\{\phi_m[k]\right\}_{k\in\kappa}$.  Gedalyahu, Tur and Eldar~\cite{Gedalyahu01} suggest a practical approach for obtaining the set $\left\{\phi_m[k]\right\}_{k\in\kappa}$, involving a bank of modulators and integrators.  Referring to Figure~\ref{Fig:02}, after having chosen the set of indices $\kappa$, we construct $p\geq K$ branches, each comprising a modulating kernel and an integrator.  We then set the modulating kernels to be: 
\begin{equation}\label{E:09}
\begin{array}{cc}
s_q\left(t\right)=\sum_{k\in \kappa}{\left(s_{q,k}e^{-j\frac{2\pi}{\tau}kt}\right)},
& q=1,2,...,p \end{array}.
\end{equation}
\noindent Ref.~\citenum{Gedalyahu01} proves that the following relation holds: 
\begin{equation}\label{E:10}
\bf{c=S\bf{\phi}}, 
\end{equation} 
where ${\bf{S}}$ is a $p\times K$ matrix with $s_{q,k}$ as its $(q,k)$ element, ${\bf{c}}$ denotes the length-$p$ sample vector with the output of the $q$th branch as its $q$th element, and ${\bf{\phi}}$ denotes the length-$K$ vector with the Fourier coefficient $\phi_m\left[k\right]$ as its $k$th element. As long as ${\bf{S}}$ has full column rank, we can recover ${\bf{\phi}}$ from the samples by ${\bf{\phi}=S^{\dag}c}$.  

\noindent Denote by ${\bf{H}}$ the $K\times K$ diagonal matrix with $k$th entry $H(\omega=\frac{2\pi}{\tau}k)$, $k\in\kappa$, and by ${\bf{V\left(t\right)}}$ the $K\times L$ matrix with $(k, l)$ element
$e^{-j2\frac{2\pi}{\tau}kt_{l,m}}$, where ${\bf{t}}=\left\{t_{1,m},...,t_{L,m}\right\}$ is the vector of unknown pulse delays received in the individual transducer element.  In addition denote by ${\bf{a}}$ the length-$L$ vector whose $l$th element is $a_{l,m}$.  Then:  
\begin{equation}\label{E:11}
\bf{\phi}=HV\left(t\right)a.
\end{equation}

\noindent The matrix ${\bf{H}}$ is invertible by construction.  Generating the length-$K$ vector ${\bf{y}}$ by left multiplying ${\bf{\phi}}$ by ${\bf{H^{-1}}}$, we have:
\begin{equation}\label{E:12}
{\bf{y}=V\left(t\right)a},
\end{equation}
which is a standard problem of finding frequencies and amplitudes of a sum of $L$ cisoids (complex sinusoids). The time-delays $\left\{t_{l,m}\right\}_{l=1}^L$ may be estimated using nonlinear techniques (e.g. annihilating filter~\cite{Stoica01}, or matrix pencil~\cite{tapan01} methods).  Having obtained the time delays, estimating the amplitudes $\left\{a_{l,m}\right\}_{l=1}^L$ is a linear problem, which may be easily solved using a least squares approach.

\noindent In the next section we apply the scheme of Figure~\ref{Fig:02} on the beamformed signal $\Phi(t;\alpha=0)$, in order to reconstruct it from a small number of its samples.  Recall that standard ultrasound devices {\bf{digitally}} construct $\Phi(t;\alpha=0)$, after sampling the individual signals received in the transducer elements at the Nyquist rate.  Since our goal is to break the Nyquist barrier, we bypass the actual construction of $\Phi(t;\alpha=0)$, by translating its Xampling to a scheme which may be applied directly on the analog signals $\left\{\varphi_m(t)\right\}_{m=-M}^M$.           
\section{Generating a 2D Image by Xampling the Beam Formed Signal} 
\label{sec:04}    
\noindent At the basis of our approach is the assumption that the beamformed signal $\Phi(t;\alpha=0)$ maintains the FRI property which characterizes the signals from which it is constructed.  This property was formulated in~(\ref{E:08}).  More explicitly, $\Phi(t;\alpha=0)$ may be written in the following manner:
\begin{equation}\label{E:13}
\Phi(t;\alpha=0)=\sum_{l=1}^L{b_l h(t-t_l )}.
\end{equation}
\noindent This claim requires justification, which will not be provided within the scope of this paper.  The justification is concealed within the fact that the nonlinear scaling of $\varphi_m\left(t\right)$, formulated in~(\ref{E:07}), has little affect on the shape of the pulses which construct it, due to their extremely short support with respect to $\tau$.  

\begin{figure}
   \begin{center}
   \begin{tabular}{ll}
   \includegraphics[height=5cm]{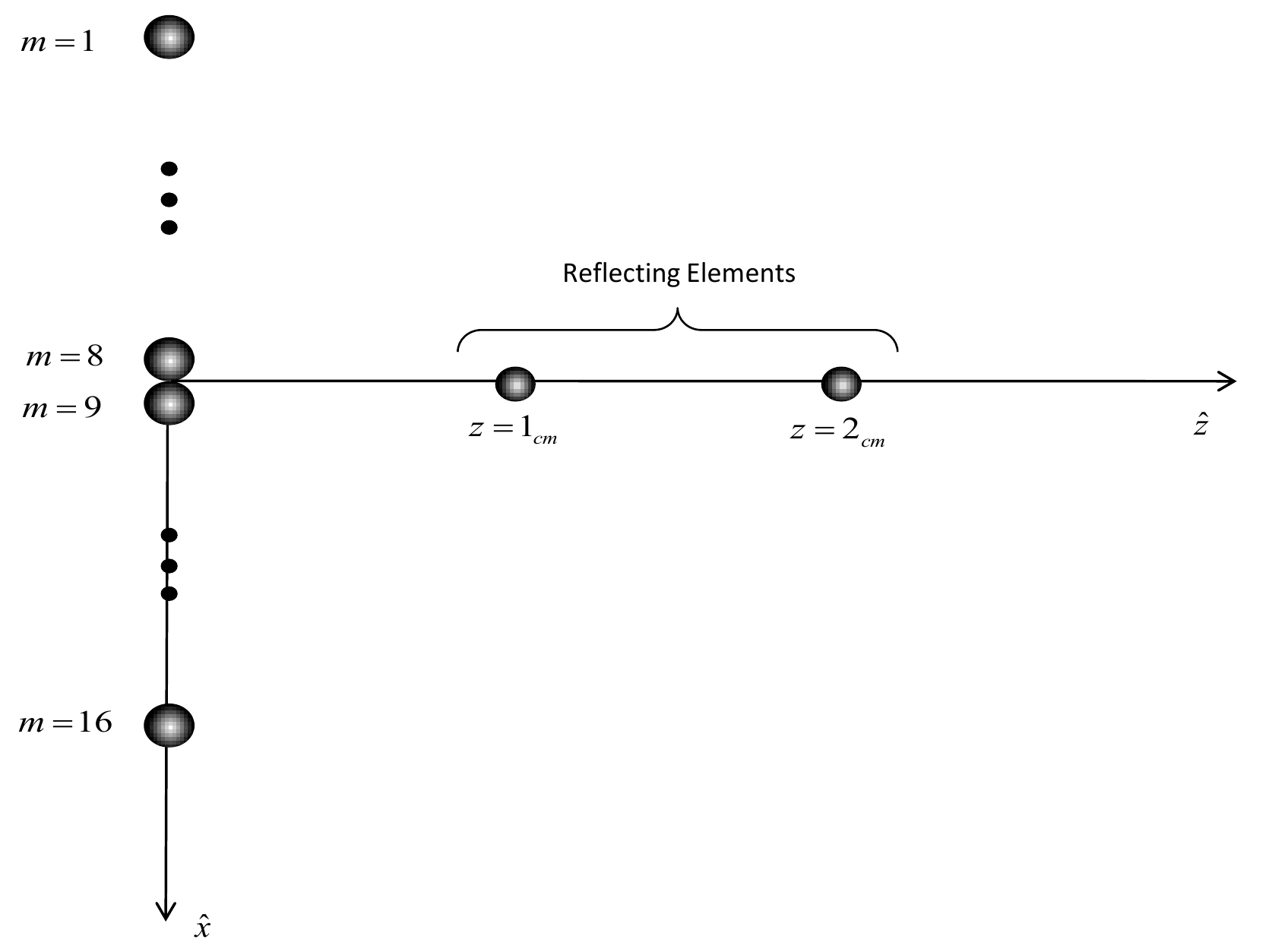} 
   \includegraphics[height=5cm]{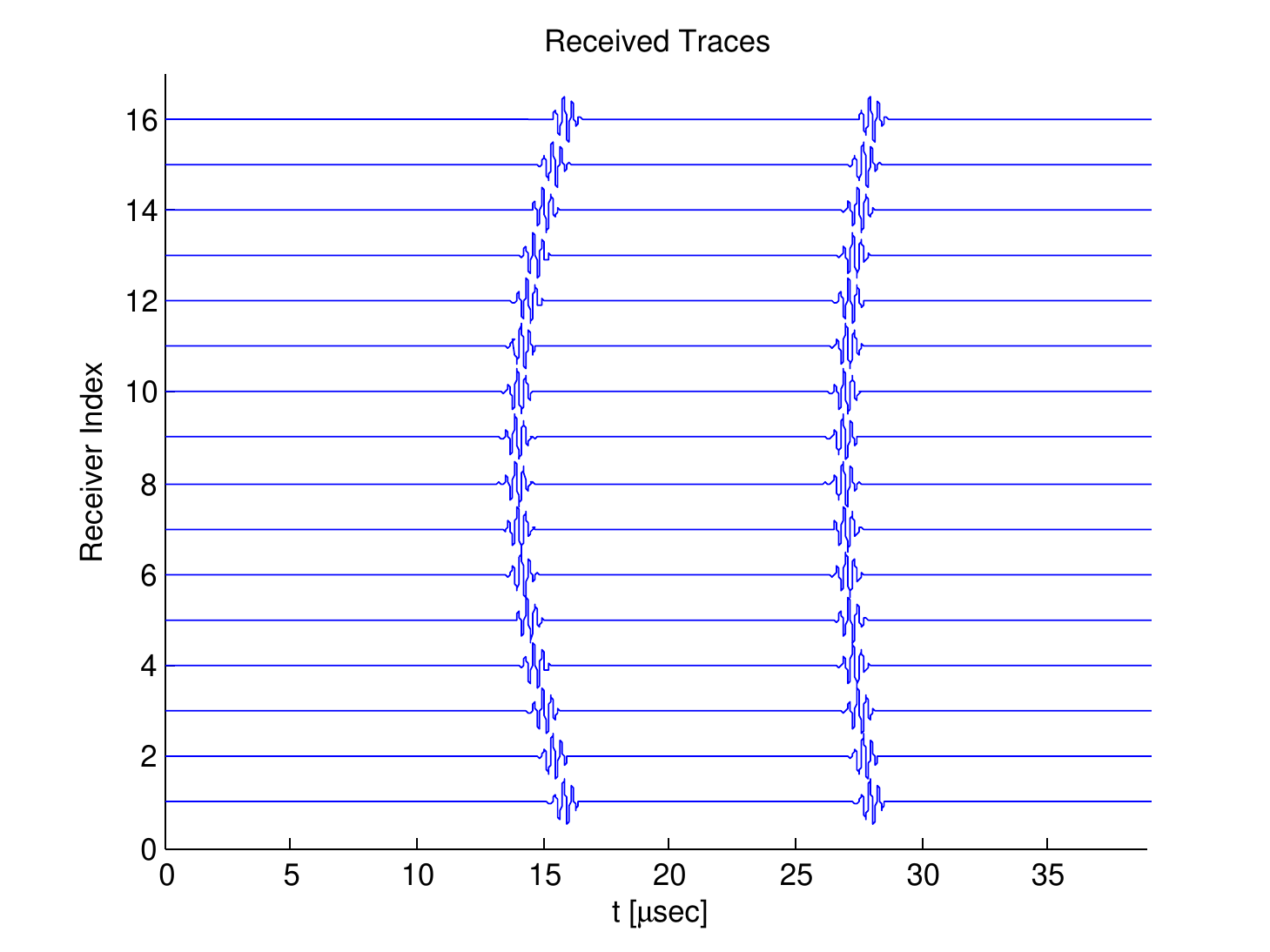} \\
   \includegraphics[height=5cm]{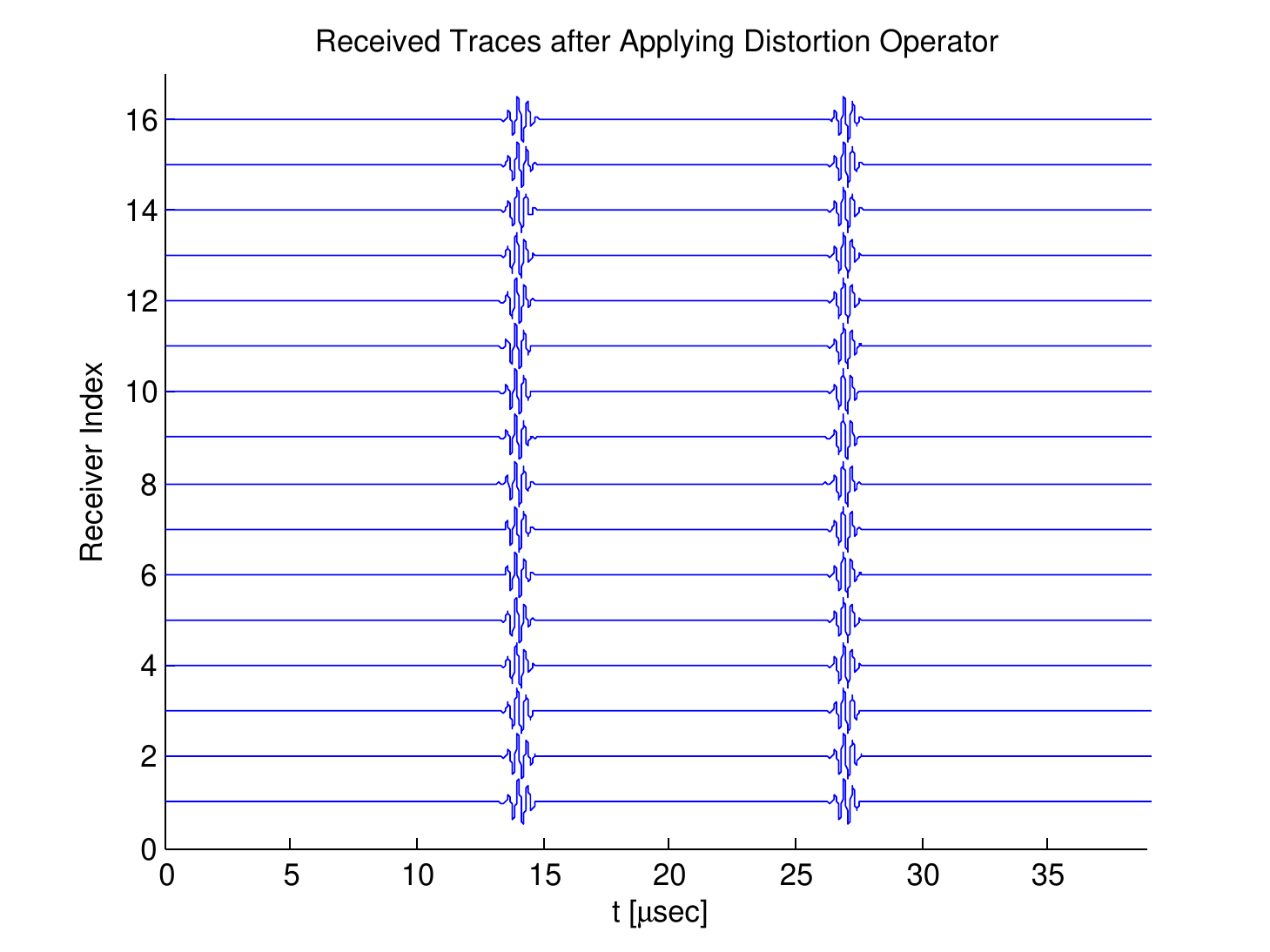} 
   \includegraphics[height=5cm]{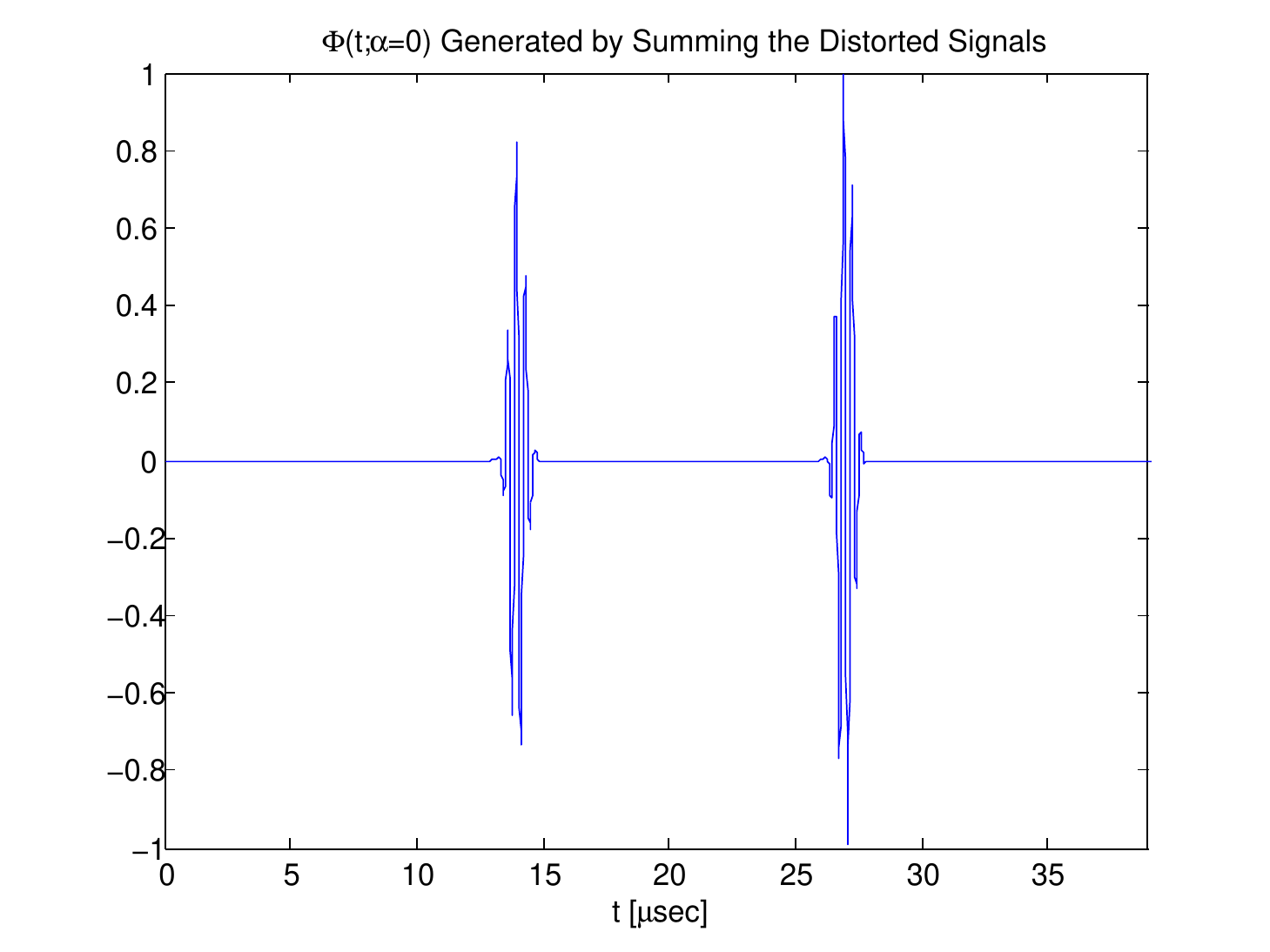} \\
   \end{tabular}
   \end{center}
   \caption[03]
   {\label{Fig:03} 
Generation of $\Phi(t;\alpha=0)$ as defined in~(\ref{E:07}), by summing distorted versions of the signals received by 16 elements.  The setup is depicted in the top left: two scattering elements are positioned along the ${\bf{\hat{z}}}$ axis, reflecting the pulse transmitted by an array of 16 elements.  The 16 traces received by the elements are depicted in the top right.  Notice that the reflected pulses are not aligned due to the different path traveled to each element.  After applying the distortion suggested in~(\ref{E:07}), we obtain modified traces, in which the pulses appear aligned, at the cost of slight distortion to the shape of each replica (bottom left).  Finally, summing the distorted traces yields the signal $\Phi\left(t;\alpha=0\right)$ (bottom right).  The latter may be approximated as a delayed sum of two known-shape pulses.}        
   \end{figure}

\noindent This notion is well demonstrated in Figure~\ref{Fig:03}.  A single pulse is transmitted along the narrow beam extending from the origin, along the ${\bf{\hat{z}}}$ axis.  Two scattering elements are illuminated by the beam (one is distanced $1_{cm}$ from the origin and the second is distanced $2_{cm}$ from the origin).  As the pulse interacts with the elements, reflections are scattered and received in each of the 16 array elements.  The traces well demonstrate the different pulse delays obtained in each of the elements.  Simply summing the 16 traces (namely - beamforming with receive focus set to infinity) may yield certain SNR improvement, yet artifacts will be formed due to the fact that the corresponding pulses are not aligned.  Using standard imaging techniques, these artifacts are manifested in a non-focused image.  Yet they may have even more profound  implications when attempting to Xample the beamformed signal.  Let us now distort each of the received traces, parametrized by $m$, as defined in~(\ref{E:07}):
\begin{equation}\label{E:14}
{\hat{\varphi}}_m(t)={\varphi}_m\left(\frac{1}{2}\left(t+\sqrt{t^2+4\left(\delta_m/c\right)^2}\right)\right).
\end{equation}
\noindent Observing Figure~\ref{Fig:03}, one may notice that the pulses are now aligned, although each may have undergone a slight distortion.  Summing the distorted signals, we now obtain $\Phi\left(t;\alpha=0\right)$ which may evidently be regarded as FRI.  If $\Phi\left(t;\alpha=0\right)$ actually existed in the analog domain, then we could Xample it using the scheme described in Section \ref{sec:03}, since it approximately satisfies~(\ref{E:13}).  Namely, we could reconstruct it from a rather small subset of Fourier coefficients corresponding to its $\tau$-periodic extension.  

\noindent Let us now assume that $\Phi\left(t;\alpha=0\right)$ existed in the analog domain, and feed it to the input of the scheme depicted in Figure~\ref{Fig:02}, instead of the individual trace $\varphi_m(t)$.  Formulating the operation of a single branch, with $\Phi(t;\alpha=0)$ as its input, we have:  
\begin{equation}\label{E:15}
 c_q=\frac{1}{\tau}{\int_0^{\tau}{\left\{\sum_{k\in \kappa}{\left(s_{q,k}e^{-j\frac{2\pi}{\tau}kt}\right)}\right\} \left\{\sum_{m=-M}^M{{{\varphi}_m\left(\frac{1}{2}\left(t+\sqrt{t^2+4\left(\delta_m/c\right)^2}\right)\right)}}\right\}dt}}.
\end{equation}
\noindent After several algebraic manipulations, and exchange of variables inside the integral,~(\ref{E:15}) may be brought into a rather straightforward form:  
\begin{equation}\label{E:16}
 c_q=\sum_{m=-M}^M\left\{{\frac{1}{\tau}{\int_0^{\hat{\tau}}{\hat{s}_{q,m}(t) {{\varphi}_m\left(t\right)}}dt}}\right\},
\end{equation}
where $\hat{s}_{q,m}(t)$ is defined as: 
\begin{equation}\label{E:17}
{\hat{s}}_{q,m}(t)\triangleq\left[1+(\frac{{\delta}_m}{ct})^2\right]\left\{\sum_{k\in\kappa}{s_{q,k}e^{-j\frac{2\pi}{\tau}k (t-\frac{1}{t}(\frac{{\delta}_m}{c})^2 )}} \right\} u(t-|\frac{{\delta}_m}{c}| ), 
\end{equation}
\noindent for $1 \leq q \leq p$, $-M \leq m \leq M$, and $\mbox{u}\left(\cdot\right)$ is the unit step function:     
\begin{equation}\label{E:18}
\mbox{u}(x)=\left\{\begin{array}{ll}1&x\geq 0 \\ 
0& \mbox{else}\end{array}\right. . 
\end{equation}  
\noindent The formulation of (\ref{E:16}) may now be interpreted by the following algorithm, applied on the set $\left\{\varphi_m(t)\right\}_{m=-M}^M$:              
\begin{enumerate}
\item Having defined the full column-rank $p\times K$ matrix ${\bf{S}}$, generate the extended set of $p\times \left(2M+1\right)$ modulating kernels ${\hat{s}}_{q,m}(t)$, defined in (\ref{E:17}).
\item Modulate each of the analog signals received by the active elements $\varphi_m\left(t\right)$, $m\in\left\{-M,...,M\right\}$, using its corresponding, size $p$, set of kernels, $\left\{\hat{s}_{q,m}\left(t\right)\right\}_{q=1}^p$, yielding a corresponding size $p$ set of coefficients:
\begin{equation}\label{E:19}
\begin{array}{cc} c_{q,m}=\frac{1}{\tau}\int_{0}^{\hat{\tau}}{{\hat{s}}_{q,m}(t){\varphi}_m(t)dt} & q\in\left\{1,...,p\right\} \end{array} .
\end{equation}   
\noindent Applying the above step upon each of the signals $\varphi_m\left(t\right)$, $m\in\left\{-M,...,M\right\}$ yields a   $p\times \left(2M+1\right)$ matrix of output samples.  
\noindent Note that the upper integration bound was modified to $\hat{\tau}$, where: 
\begin{equation}\label{E:20}
\hat{\tau}=\max_{m\in\left\{-M,...,M\right\}}\left\{{\frac{1}{2}(\tau+\sqrt{{\tau}^2+4({\delta_m/c})^2})} \right\} .
\end{equation}
\item Sum the samples along $m$, yielding the single, length-$p$ vector, ${\bf{c}}$, of which the $q$th element is $c_q$, satisfying: 
\begin{equation}\label{E:21}
c_q=\sum_{m=-M}^M{c}_{q,m} .
\end{equation} 
\item Obtain the vector ${\bf{y}}$ by: 
\begin{equation}\label{E:22}
\bf{y=H^{-1}\left(S^{\dag}c\right)} .
\end{equation}
\noindent Since we began the derivation by injecting $\Phi(t;\alpha=0)$ into the scheme depicted in Figure~\ref{Fig:02}, then the vector recovered by ${\bf{S^{\dag}c}}$, holds, as its $k$th element, the $k$th Fourier coefficient corresponding to the $\tau$-periodic extension of the {\bf{beamformed signal}} $\Phi(t;\alpha=0)$.
\item Solve the problem formulated in (\ref{E:12}), extracting the $2L$ unknowns of  (\ref{E:13}).  Here ${\bf{V\left(t\right)}}$ is the $K\times L$ matrix with $(k,l)$ element $e^{-j2\frac{2\pi}{\tau}kt_{l}}$, ${\bf{t}}=\left\{t_{1},...,t_{L}\right\}$ is the vector of unknown pulse delays, and ${\bf{a}}$ is the length-$L$ vector whose $l$th element is $b_{l}$.   The time delays are extracted by applying nonlinear techniques (e.g. annihilating filter, or matrix pencil methods).  Estimating the amplitudes $\left\{b_{l}\right\}_{l=1}^L$ is then a linear problem.  
\end{enumerate}

\begin{figure}[H]
   \begin{center}
   \begin{tabular}{c}
   \includegraphics[height=18cm]{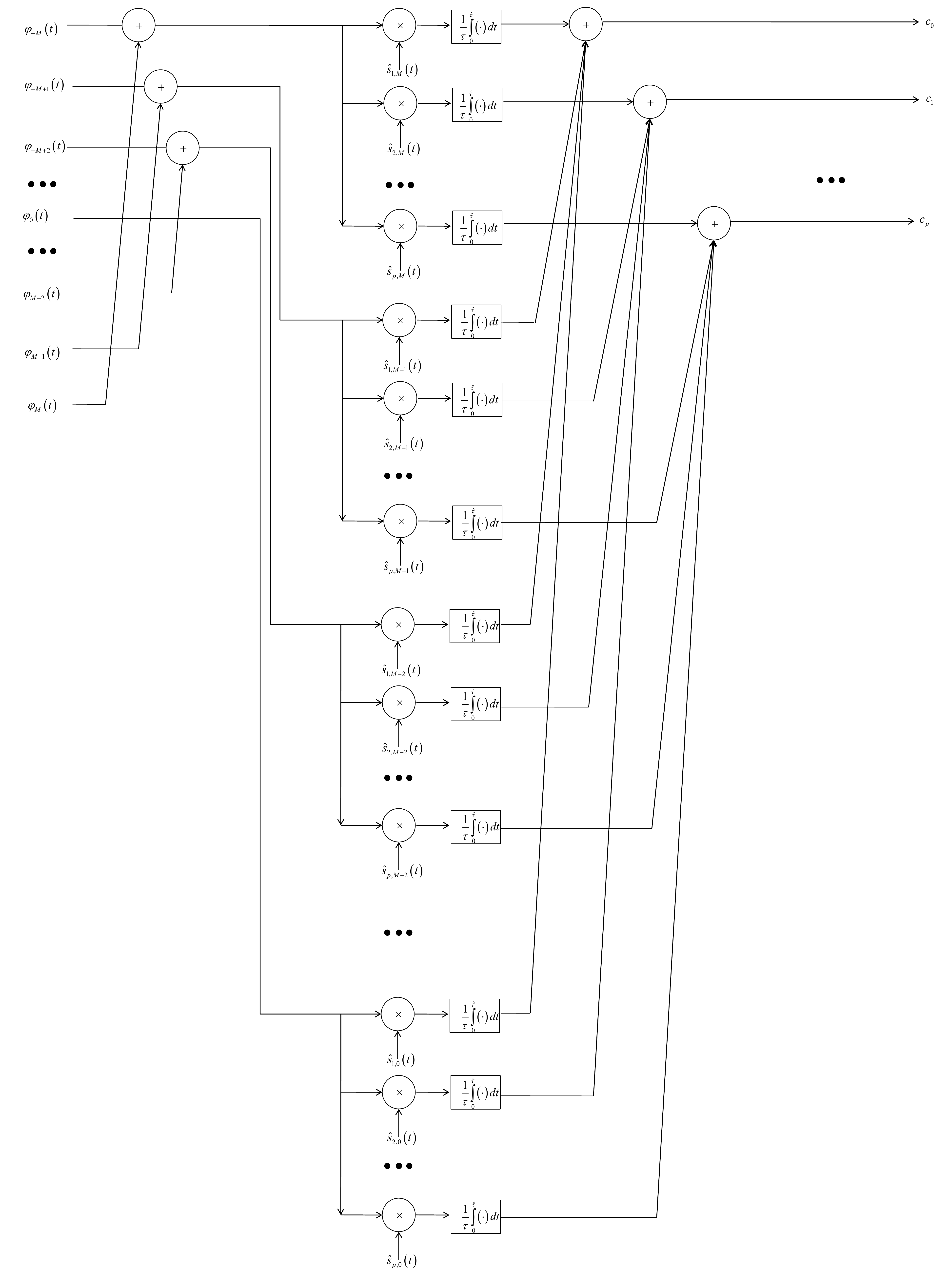}
   \end{tabular}
   \end{center}
   \caption[04]
   {\label{Fig:04} Generalized Xampling scheme yielding $p$ coefficients, used for reconstructing $\Phi(t;\alpha=0)$ from $2M+1$ receiving elements.}    
   \end{figure}

\noindent Summarizing the above,  Figure~\ref{Fig:04} schematically depicts the suggested generalized Xampling scheme.  The active array elements typically lie symmetrically with respect to the processed image line, resulting in symmetric kernels, with respect to $m$ (i.e. $\hat{s}_{q,-m}\left(t\right)=\hat{s}_{q,m}\left(t\right)$).  Exploiting this symmetry enables reducing the total number of samples, by summing each pair $\varphi_m(t)$ and $\varphi_{-m}(t)$ prior to the modulation and integration branch. This concept is implemented in the figure.  

\noindent Recall that the transmitted (and reflected) pulses are modulated to a high frequency carrier (typically $f_c=5_{MHz}$), such that most of the pulse energy is concentrated far from the DC frequency.  In order to avoid singularity of the matrix ${\bf{H}}$ it is necessary to pick the set $\kappa$, such that the frequencies $\left\{\omega_k=\frac{2\pi} {\tau}k\right\}_{k\in\kappa}$ are near $2\pi f_c$.  This implies, that when implementing the Xampling scheme depicted in Figure~\ref{Fig:02} using real filters, the set $\kappa$ is doubled, so that if the index $k$ is in the set $\kappa$, we must also add $-k$ to $\kappa$.  With the condition that $p\geq |\kappa|$ (so that ${\bf{S}}$ is full column rank), $|\cdot|$ denoting cardinality, we are forced to double the number of samples:  for $\Phi(t=0;\alpha)$ having $2L$ degrees of freedom ($\left\{b_l,t_l\right\}_{l=1}^L$), reconstruction requires a minimum of $2K$ samples within the interval $\left[0,\tau\right)$, where $K\geq2L$ as before.  Indeed, the actual ultrasound signal Xampled in Ref.~\citenum{Tur01} was first demodulated, yielding a ``complex'' analog signal.  This practically means that the single, modulated real signal, was split into two base-band signals, each Xampled at least at the Rate of Innovation - thereby the number of samples obtained within the interval $\left[0,\tau\right)$ satisfied $p\geq2K\geq4L$.

\noindent To conclude, we Xample the signals $\left\{\varphi_m(t)\right\}_{m=-M}^M$ in a manner which enables reconstruction of the beamformed signal $\Phi(t;\alpha=0)$, from which an image line may be directly generated: the individual delay $t_l$ corresponds to a coordinate, located along the line normal to the set of active elements, at a distance $\frac{ct_l}{2}$ from its center.  We may simply set the pixel associated with this coordinate an intensity proportional to $b_l$.   Alternatively, we may generate a trace, by convolving the stream of pulses $\sum_{l=1}^L{b_l\delta(t-t_l)}$ with the envelope of $h(t)$, and then set the pixel intensities along the corresponding image line accordingly.  The latter method was used for the simulations discussed in the next section, in order to obtain results comparable to those obtained by standard imaging techniques.  

\noindent For our simulations, we assumed $L=30$ macroscopic scatterers along a single beam.  Defining $\rho\geq 1$ to be the oversampling factor, we then need at least $K=2\rho L$ consecutive indices, for which $H(\omega=\frac{2\pi}{\tau}k)\neq 0$.  We select the set of indices, $\kappa$, such that $\omega_k=\frac{2\pi}{\tau}k$ are near $2\pi f_c$, where $f_c$ is the frequency of the carrier wave (approximately $5_{MHz}$).  In order to obtain real kernels, we add the opposite indices to the set $\kappa$, such that we finally have $|\kappa|=4\rho L$.  We now construct a scheme with $p$ channels per  element, $-M\leq m \leq M$, such that $p=|\kappa|=4\rho L$.  We do this by choosing the matrix ${\bf{S}}$ in a very straightforward manner:  
\begin{equation}\label{E:23}
{\bf{S}}_{p\times p}=\left[\begin{array}{cc} \frac{1}{2}{\bf{I}} & \frac{1}{2}{\bf{I}}\\ \frac{1}{2j}{\bf{I}} & -\frac{1}{2j}{\bf{I}}\end{array}\right], 
\end{equation} 
where ${\bf{I}}$ is a $\frac{p}{2}\times\frac{p}{2}$ identity matrix.  Referring to~(\ref{E:17}), our selection of ${\bf{S}}$ forms the following set of $p\times (2M+1)$ kernels:
\begin{equation}\label{E:24} 
\hat{s}_{q,m}\left(t\right) = \left\{\begin{array}{cc}\left[1+(\frac{{\delta}_m}{ct})^2\right]\mbox{cos}\left(\frac{2\pi}{\tau}k_q \left(t-\frac{1}{t}(\frac{{\delta}_m}{c})^2 \right)\right)u\left(t-|\frac{{\delta}_m}{c}|\right) & 1\leq q \leq \frac{p}{2}\\
-\left[1+(\frac{{\delta}_m}{ct})^2\right]\mbox{sin}\left(\frac{2\pi}{\tau}k_{q-\frac{p}{2}} \left(t-\frac{1}{t}(\frac{{\delta}_m}{c})^2 \right)\right)u\left(t-|\frac{{\delta}_m}{c}|\right) & \frac{p}{2}+1\leq q \leq p \\
\end{array}\right. ,
\end{equation}       
\noindent which were utilized in the scheme depicted in  Figure~\ref{Fig:04}.  

\noindent We chose the matrix pencil~\cite{tapan01}  method in order to estimate the set of delays from the estimated Fourier coefficients obtained using our scheme.  One significant advantage of the matrix pencil scheme is that it provides the ability to estimate the actual number of delays concealed within the noisy data, based on a SVD decomposition process.  Recall that in real imaging we have no prior knowledge regarding the number of reflecting elements aligned along a single image line.  Nevertheless, when we design the matrix pencil we must determine the pencil parameter, which we shall denote by $\eta$.  The matrix pencil method assumes $\eta$ to be greater than (or equal to) the number of complex exponentials comprising the estimated signal (in our context, this is equivalent to the number of delayed pulses, $L$).  As a result, when designing the matrix pencil block we must first determine an upper bound on the number of reflected elements along a single image line and then set the pencil parameter $\eta$ accordingly.  This farther dictates a lower bound on the number of samples used for estimating the delays (denoted here by $K$).  Summarizing the above constraints, the following must hold:
\begin{equation}\label{E:25}
L\leq \eta \leq K-L, 
\end{equation} 
\noindent where $L$ is an estimated upper bound on the number of reflecting elements.  The need to guarantee a non-empty interval $\left[L, K-L\right]$, justifies the requirement that  $K\geq 2L$.    
\section{Results} 
\label{sec:05}  
\noindent In the following section, we examine the result of applying our suggested Xampling scheme upon actual raw RF ultrasound data.  The data was acquired using a programmable imaging system (Model V-1-128, Verasonics, Inc., Redmond, WA), equipped with a 128-element 1-D linear transducer array (Model L7-4, Philips Healthcare, Bothell, WA). The imaging target was a commercial multi-purpose gray-scale phantom (Model 403GS LE, Gammex, Inc., Middleton, WI) including 0.1-mm nylon wires embedded in tissue mimicking material.  We compare the following images:
\begin{enumerate}
\item Standard Image - Ultrasound image generated using standard imaging technique:  the frequency of the carrier wave was $5.142_{MHz}$;  data was acquired  at high rate ($20_{MHz}$);  the number of active transducer elements used for focusing the beam at transmit (Tx) and at receive (Rx) was 16.  Setting the maximum imaging depth to $7.88_{cm}$, the number of samples used in order to generate a single image line was thus $16\times 2048=32,768$.  We used dynamic Rx focusing, comprising $K=100$ focal zones.     
\item Xampled Image, Dynamic Rx Focusing - Ultrasound image generated using the exact scheme suggested in Section \ref{sec:04}.      
\item Xampled Image, Infinity Rx Focusing - Ultrasound image generated using the scheme suggested in Section \ref{sec:04}, forcing $\delta_m=0$, $\forall m\in\left\{-M,...,M\right\}$.  Observing~(\ref{E:07}), one can easily see that this degenerate implementation forces Rx focusing to be infinity: $\Phi\left(t;\alpha\right)$ is  simply the sum of all signals obtained from the active elements, without applying any delay.      
\end{enumerate}

\begin{figure}[H]
   \begin{center}
   \begin{tabular}{c}
   \includegraphics[height=9cm]{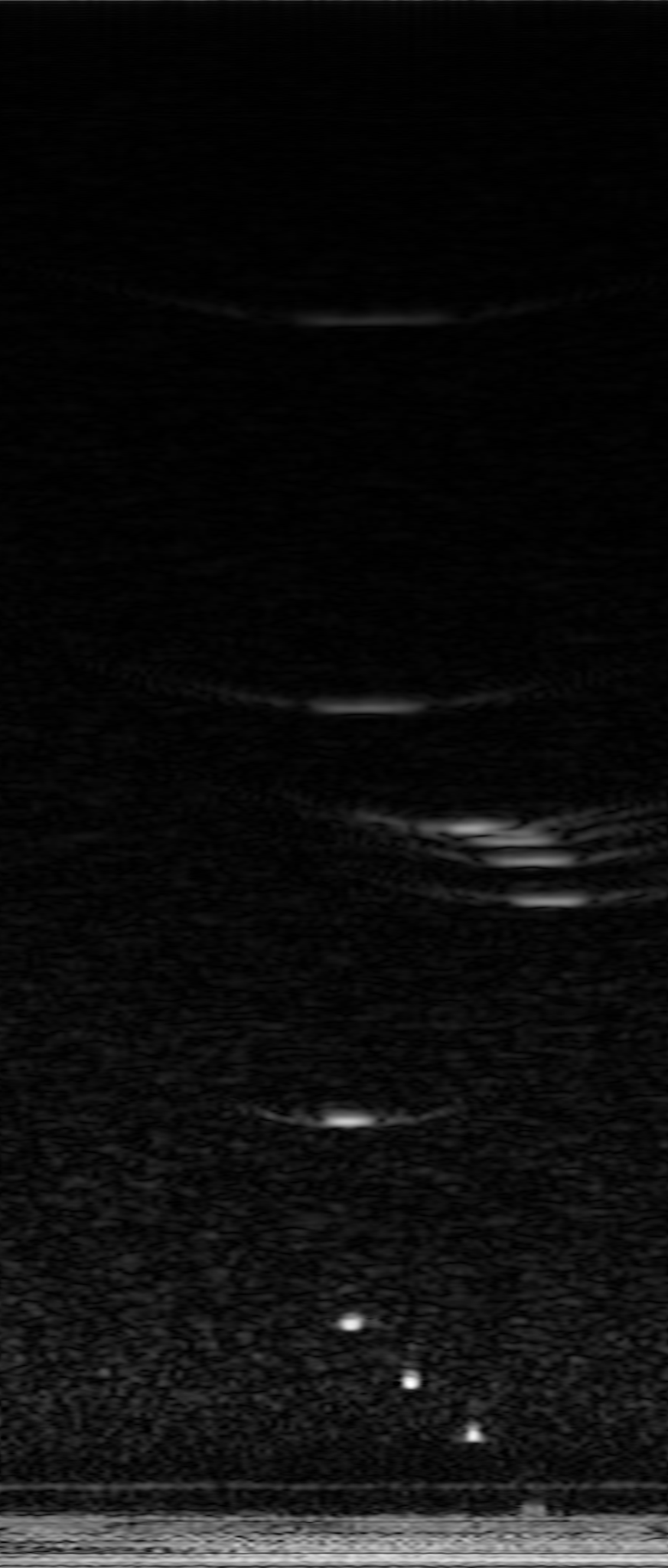}
   \end{tabular}
   \end{center}
   \caption[05]
   {\label{Fig:05} Standard image comprising 113 Image Lines, generated with 100 focal zones.}    
   \end{figure}

\begin{figure}[H]
   \begin{center}
   \begin{tabular}{cc}
   \includegraphics[height=9cm]{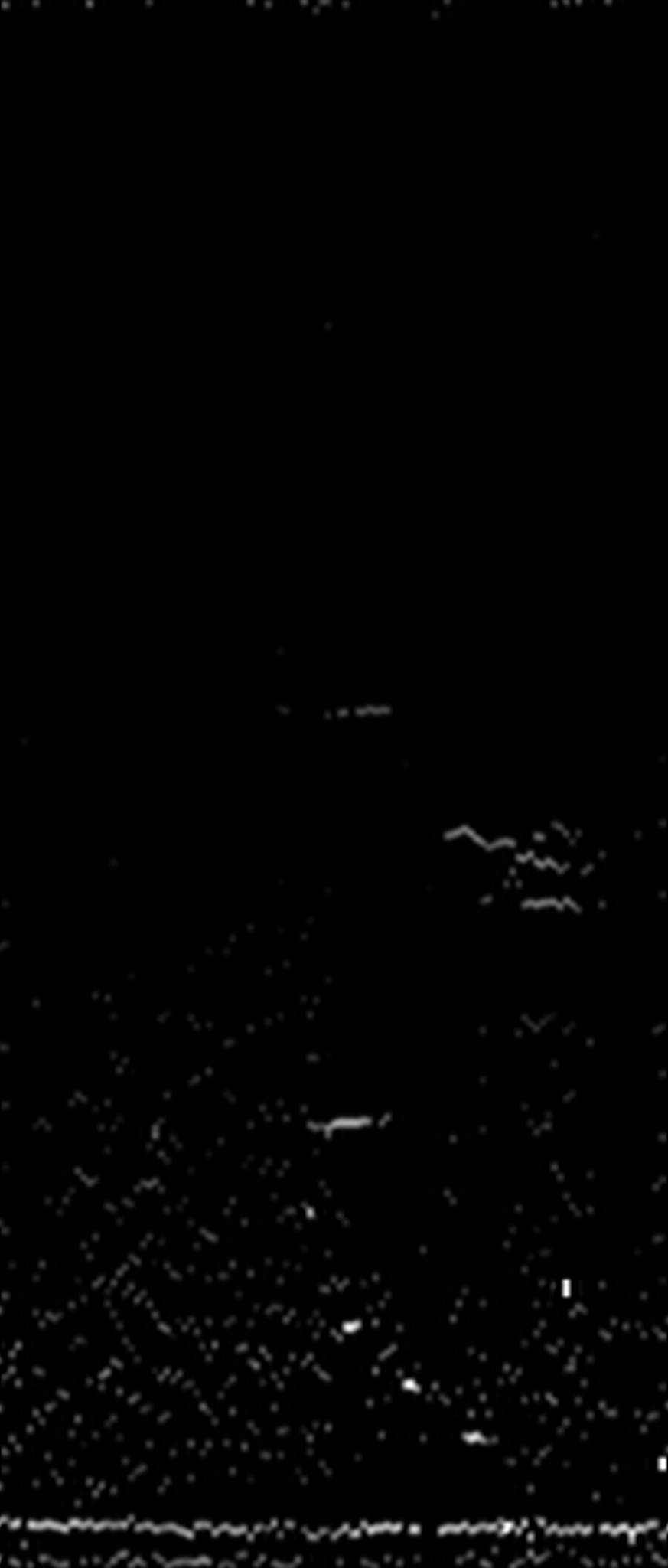}&
   \includegraphics[height=9cm]{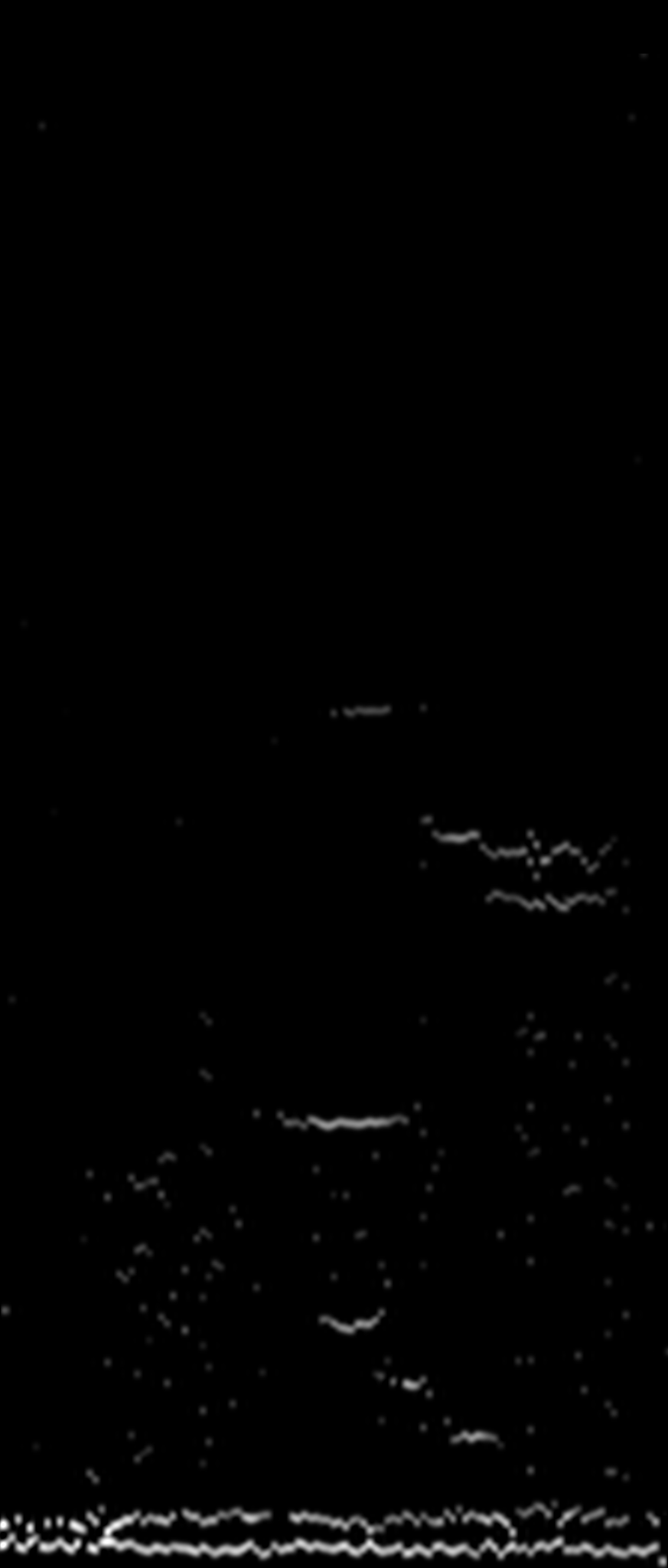}
   \end{tabular}
   \end{center}
   \caption[06]
   {\label{Fig:06} Xampled Images with $L=30$ and $\rho=1$.  Dynamic Focusing (left) and Focus at Infinity (right).}    
   \end{figure}

\begin{figure}[H]
   \begin{center}
   \begin{tabular}{cc}
   \includegraphics[height=9cm]{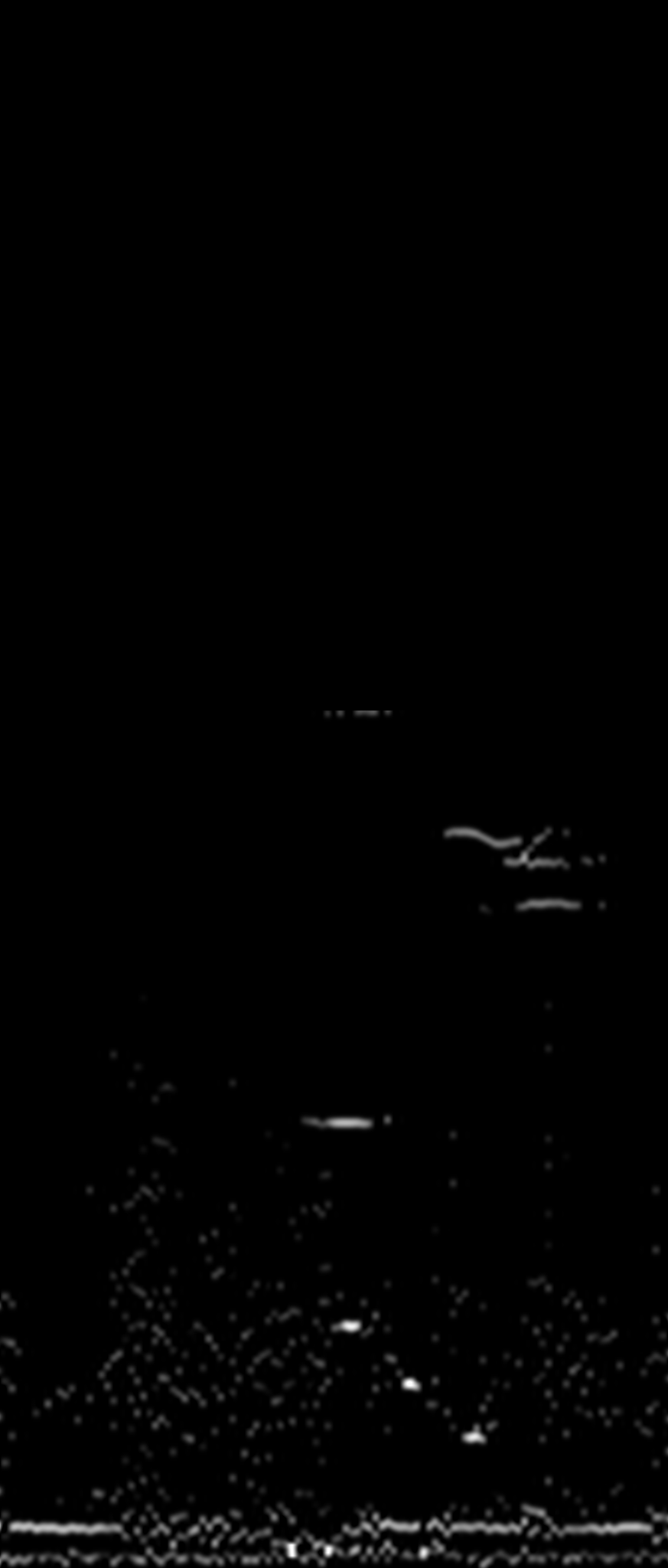}&
   \includegraphics[height=9cm]{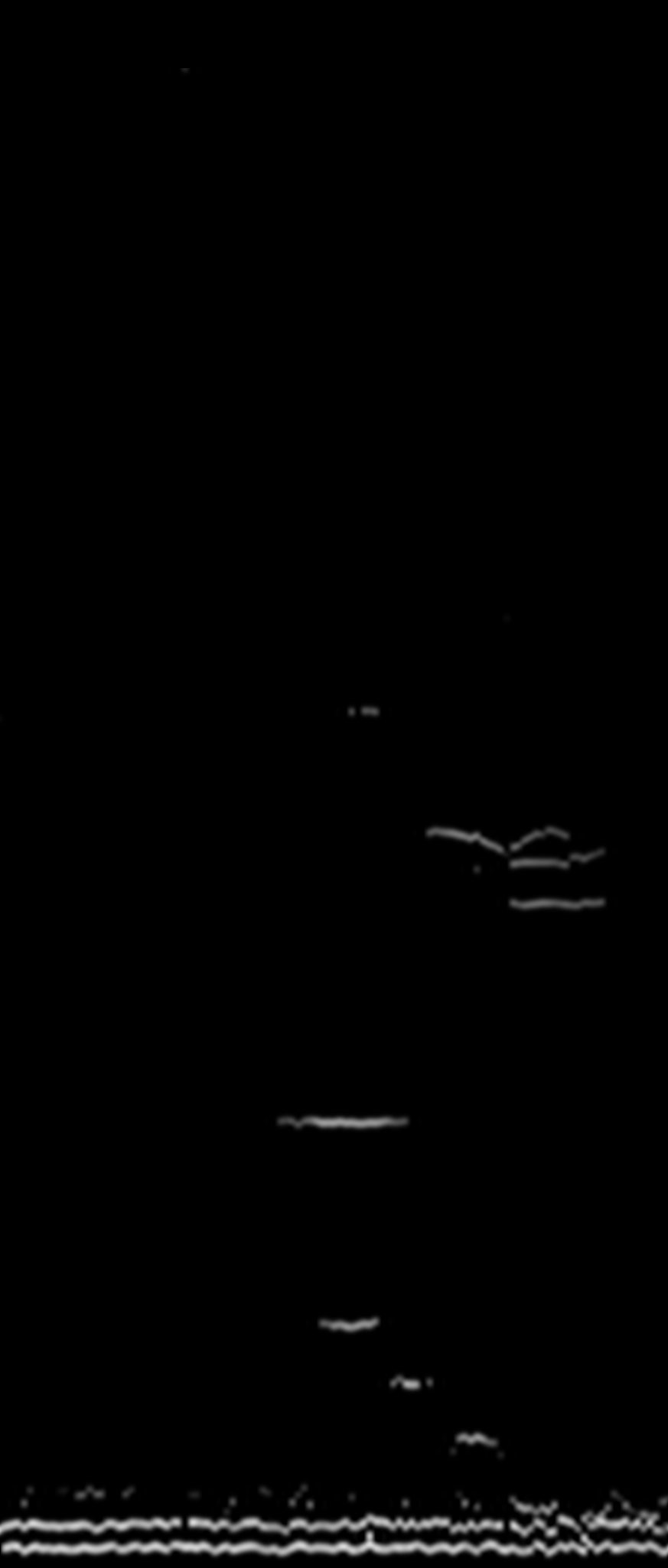}
   \end{tabular}
   \end{center}
   \caption[07]
   {\label{Fig:07} Xampled Images with $L=30$ and $\rho=2$.  Dynamic Focusing (left) and Focus at Infinity (right).}    
   \end{figure}

\begin{figure}[H]
   \begin{center}
   \begin{tabular}{cc}
   \includegraphics[height=9cm]{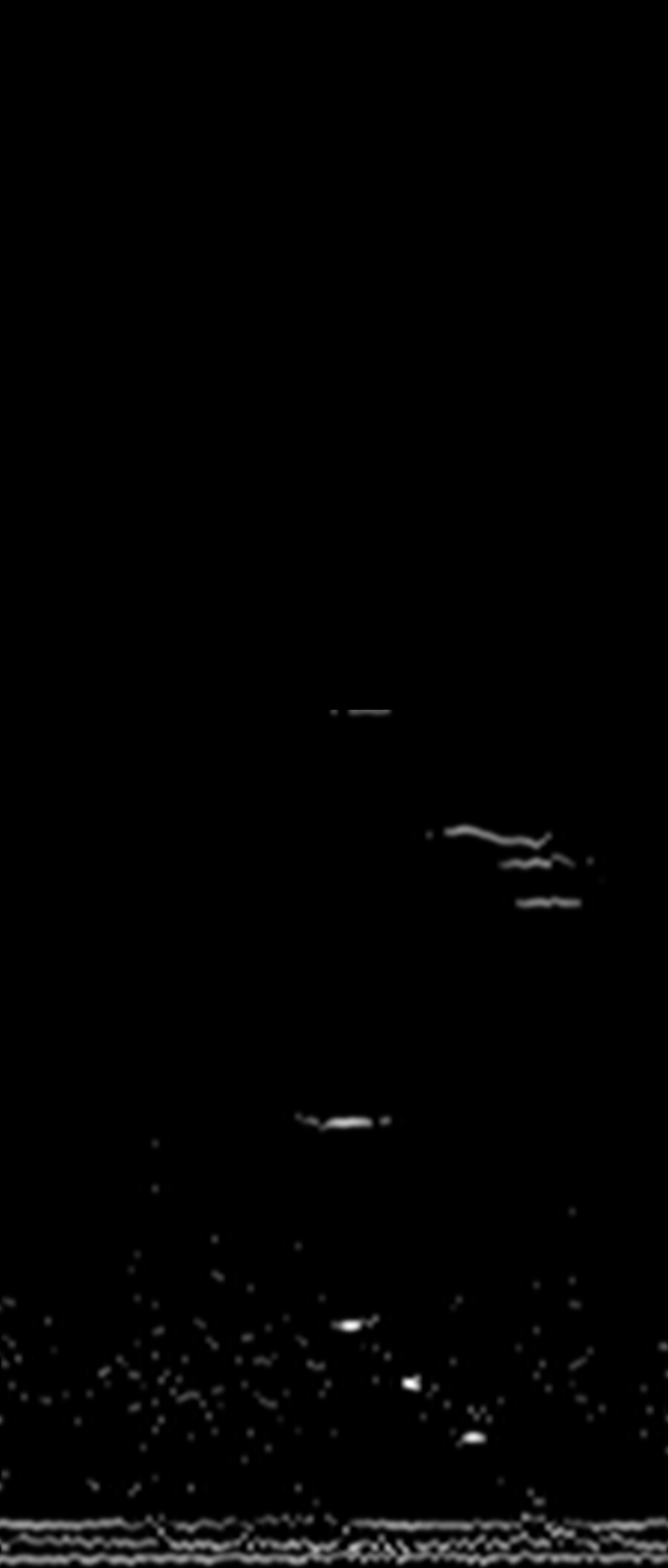}&
   \includegraphics[height=9cm]{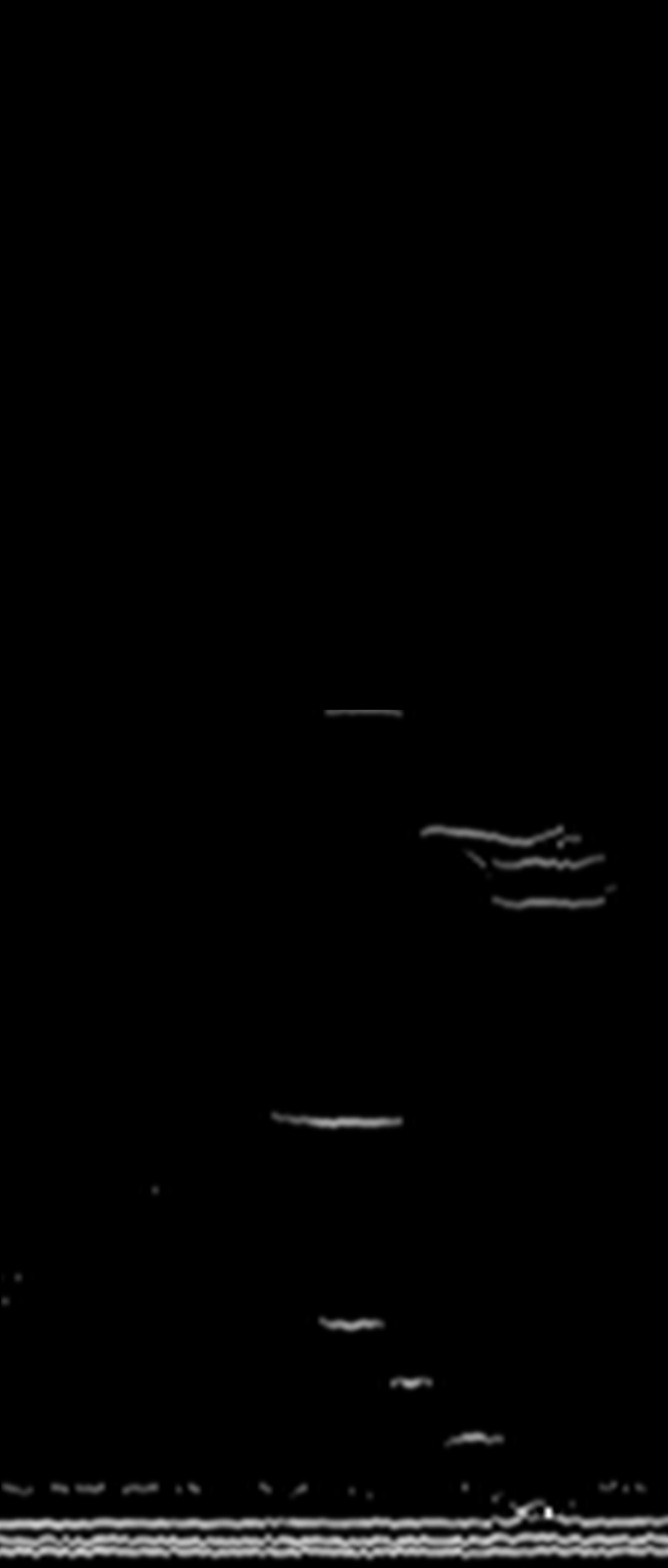}
   \end{tabular}
   \end{center}
   \caption[08]
   {\label{Fig:08} Xampled Images with $L=30$ and $\rho=3$.  Dynamic Focusing (left) and Focus at Infinity (right).}    
   \end{figure}

\begin{figure}[H]
   \begin{center}
   \begin{tabular}{cc}
   \includegraphics[height=9cm]{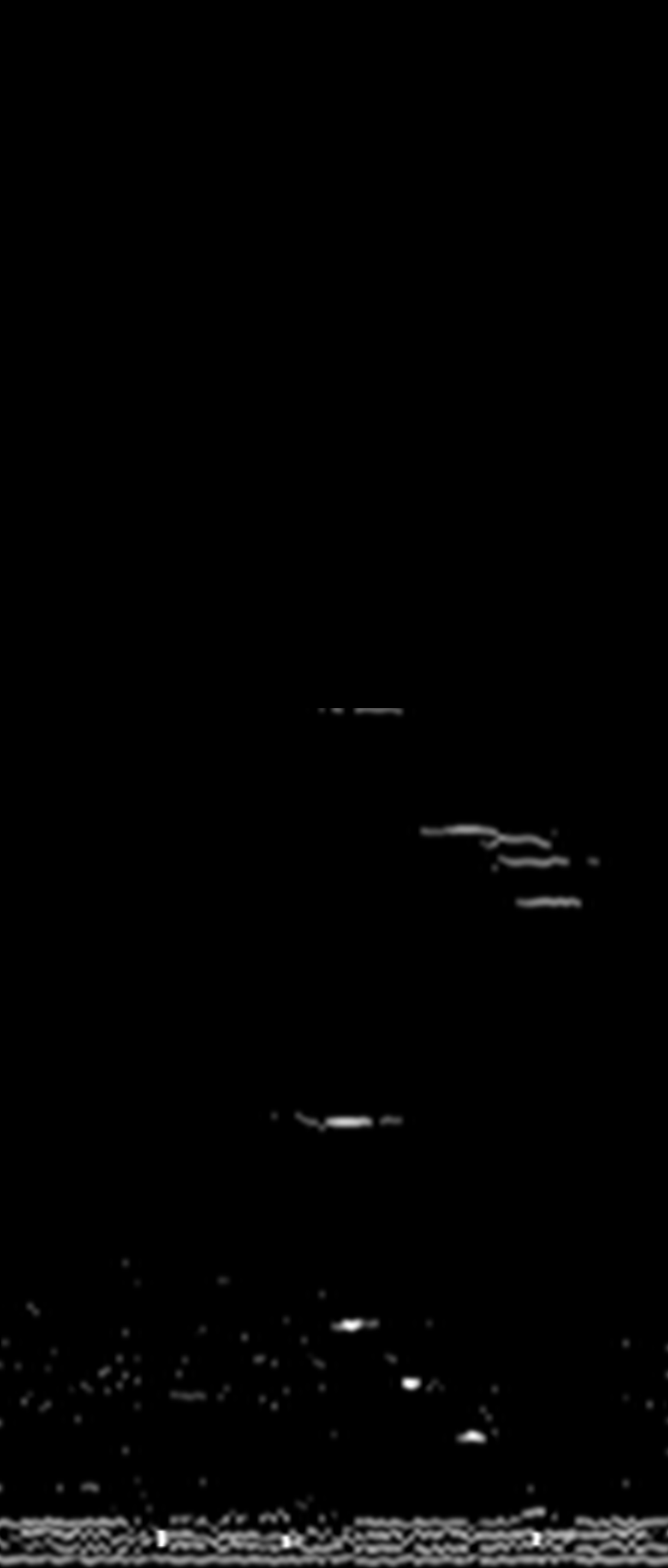}&
   \includegraphics[height=9cm]{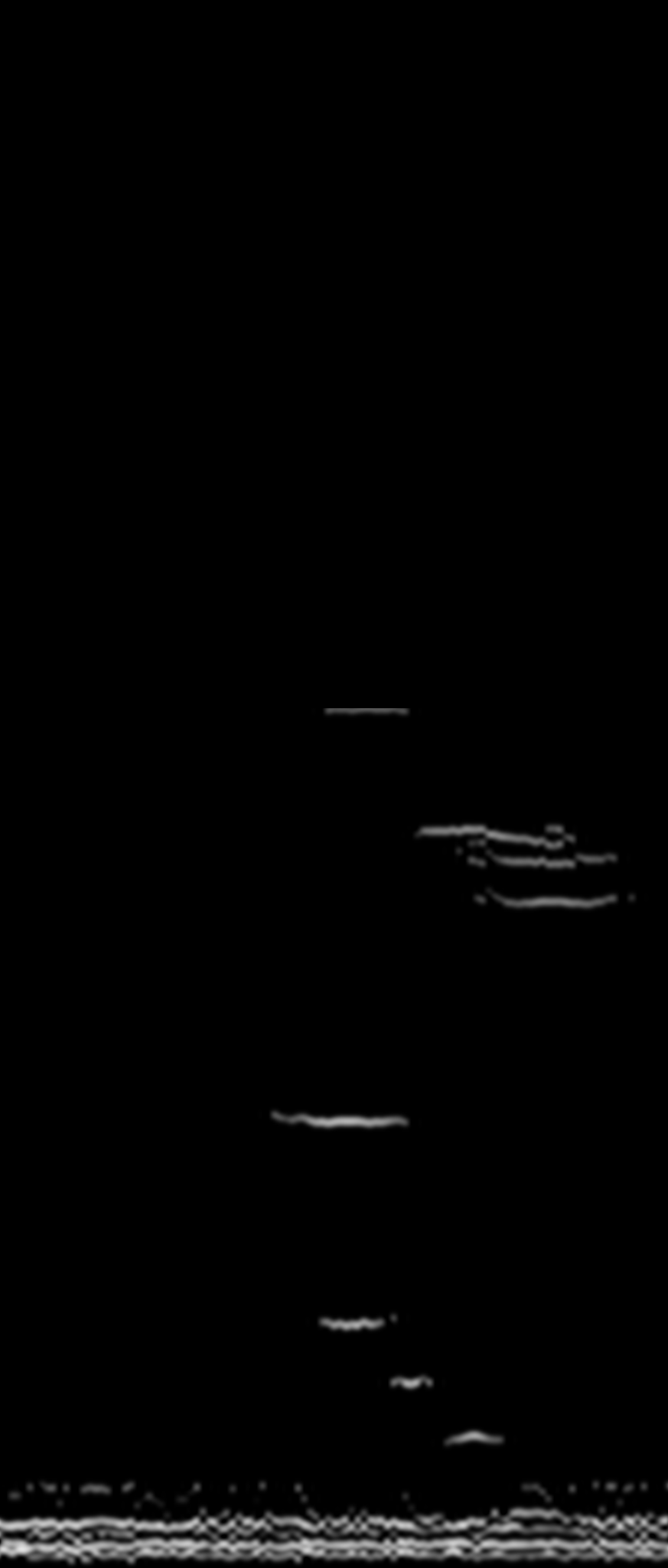}
   \end{tabular}
   \end{center}
   \caption[09]
   {\label{Fig:09} Xampled Images with $L=30$ and $\rho=4$.  Dynamic Focusing (left) and Focus at Infinity (right).}    
   \end{figure}

\noindent The resulting images are arranged in Figures~\ref{Fig:05}-\ref{Fig:09}, according to the following table:
\begin{table}[h]
\centering
\begin{tabular}{|p{0.6in}|p{2.2in}|p{1.5in}|} \hline
Figure & Image Type & Estimated No. of \\ 
No. &  & Elements/Oversampling Factor \\ \hline\hline
5  & Standard & Irrelevant \\ \cline{1-3}
6 & Xampled (Dyn. Focus and $\infty$ Focus) & 30/1 \\ \cline{1-3}
7 &  Xampled (Dyn. Focus and $\infty$ Focus) & 30/2 \\ \cline{1-3}
8 &  Xampled (Dyn. Focus and $\infty$ Focus) & 30/3 \\ \cline{1-3}
9 &  Xampled (Dyn. Focus and $\infty$ Focus) & 30/4 \\ \hline
\end{tabular}
\caption{Properties of images displayed in Figures~\ref{Fig:05}-\ref{Fig:09}}
\label{Tab:01}
\end{table}


\noindent At this point of our work we avoided deriving quantitative measures to the estimation quality;  this is because we had no actual documentation regarding the true positions and intensities of the reflecting elements.  Using SNR measurements with the ``Standard Image" as reference may be very deceiving, since the latter is obviously noisy by itself - noise which was often eliminated by the Xampling scheme, due to the fact that it extracts isolated delta excitations.  The reader may receive qualitative impressions, observing the similarity between the Xampled images and the ``Standard" image, and the obvious difference between a dynamically focused image and its equivalent infinity focused one.  

\noindent In Table \ref{Tab:02} we provide an estimate of: a) the number of samples required for generating a single image line using our suggested Xampling scheme (referring to $L$ and $\rho$ used in Figures~\ref{Fig:06}-\ref{Fig:09}) and using standard imaging techniques; b) the estimated computational cost of generating the same image line for both methods: 
  
\begin{table}[H]
\centering
\begin{tabular}{|p{0.8in}|p{0.6in}|p{0.8in}|p{0.25in}||p{1.1 in}|p{1.1 in}|} \hline
Image Type &Estimated& Oversampling& $K$ & Sampling Rate & Cost \\ 
 &No. of  & Factor & &[No. of Samples/  & [MegaOps./Line] \\ 
 &Reflectors & & & Element/Line] & \\ \hline\hline
Xampled&30  & 1 & 60 & 120 & 0.43 \\ \cline{2-6}
&30  & 2 & 120 & 240 & 2.81 \\ \cline{2-6}
&30  & 3 & 180 & 360 & 9.06 \\ \cline{2-6}
&30  & 4 & 240 & 480 & 21.05 \\ \cline{1-6}
Standard&  \multicolumn{2}{l}{~~~~~~~~~~~~~~~~Irrelevant}& &  2048 & 0.06 \\ \cline{1-6}
\end{tabular}
\caption{Sampling rates and computational costs estimated for Xampled imaging (using various over-sampling factors) and for standard imaging.  The computational cost of a Xampled image line roughly grows like $O\left(|\kappa|^3\right)$.}
\label{Tab:02}
\end{table}

\noindent The computational cost required for the standard imaging process (assuming 2048 samples per image line, and 16 active receivers used for beamforming) comprises $2048\times15$ Add operations, and additionally the cost of the Hilbert transform which is utilized for envelope detection (typically implemented using two FFT operations).   Table \ref{Tab:03} which appears in {\bf{Appendix A}} details the blocks used for estimating  the Xampling scheme computational costs.
\section{Conclusions and Future Work} 
\label{sec:06}  
\noindent This work focused on generalizing the Xampling method suggested in Ref.~\citenum{Gedalyahu01} to an array of multiple receiving elements.  At the heart of our generalization is the observation, that the dynamic focusing and the filtering part of the Xampling can be combined into a set of modulating kernels and performed directly on the analog signals. This, in turn, can be sampled at a rate way below the Nyquist rate.
   
\noindent Preliminary tests on actual ultrasound data yield results which are quite similar to an image obtained using standard techniques, while reducing the sampling rate by a factor of 5 to 15.  Apparently, this is achieved at the cost of increased computational effort.  
\noindent By reducing the sampling rate, we hope to simplify the front end hardware (mainly in terms of size and power consumption) while maintaining image quality.     

\acknowledgments     
\noindent  The authors would like to thank Dr. Omer Oralkan and Prof. Pierre Khuri-Yakub of the E. L. Ginzton Laboratory at Stanford University, for providing the RF ultrasound data and for many helpful discussions.

\appendix    
\section{Estimation of Xampling scheme computational cost }
\begin{table}[H]
\centering
\begin{tabular}{|p{1.0 in}|p{2.5in}|p{1.5 in}|} \hline
Operation & Details & Cost/$O\left(\cdot\right)$ \\ \hline\hline
Sum Outputs & $Sum\left({\bf{c}}_{L\times(2M+1)},2\right)$ & $\left(L-1\right)\times\left(2M+1\right)$\\ \hline
Matrix Pencil & ${\bf{Y}}_{K\times1 }={\bf{A}}_{K\times p}{\bf{c}}_{p\times1}$ & $K\times P$\\ \cline{2-3}
Obtain $\left\{t_l\right\}_{l=1}^L$ & $SVD\left({\bf{Q}}_{\frac{2}{3}K\times\frac{1}{3}K}\right)$ & $\frac{2}{3}K\times\left(\frac{1}
{3}K\right)^2$\\ \cline{2-3}
 & ${\bf{Y}}_{\frac{2}{3}K\times\frac{1}{3}K }={\bf{U}}_{\frac{2}{3}K\times L}{\bf{\Sigma'}}_{L\times L}{\bf{V'^T}}_{L\times \frac{1}{3}K}$ & $\frac{2}{3}K\times L \times L + L \times L \times \frac{1}{3} K$\\ \cline{2-3}
& ${\bf{Y}}_{\frac{2}{3}K\times\frac{1}{3}K }={\bf{U}}_{\frac{2}{3}K\times L}{\bf{\Sigma'}}_{L\times L}{\bf{V'^T}}_{L\times \frac{1}{3}K}$ & $\frac{2}{3}K\times L \times L + L \times L \times \frac{1}{3} K$\\ \cline{2-3}
& $pinv\left({\bf{Y}}_{\frac{2}{3}K\times\frac{1}{3}K }\right)$ & $\left(\frac{2}{3}K\right)^3+\left(\frac{1}{3} K\right)^3$\\ \cline{2-3}
& ${\bf{Q}}_{\frac{1}{3}K\times\frac{1}{3}K }={\bf{A}}_{\frac{1}{3}K\times \frac{2}{3}K}{\bf{B}}_{\frac{2}{3}K\times \frac{1}{3}K}$ & $\frac{2}{3}K\times \left(\frac{1}{3} K\right)^2$\\ \cline{2-3}
& $eig\left({\bf{Q}}_{\frac{1}{3}K\times\frac{1}{3}K }\right)$ & $ \left(\frac{1}{3} K\right)^3$\\ \hline
Least Squares & ${\bf{L}}_{K\times L }\star {\bf{M}}_{K\times L }$ & $ K\times L$\\ \cline{2-3}
Obtain $\left\{b_l\right\}_{l=1}^L$ & $pinv\left({\bf{V}}_{K\times L}\right)$ & $K^3+L^3$\\ \cline{2-3}
 & ${\bf{b}}_{L\times1 }={\bf{A}}_{L\times K}{\bf{c}}_{K\times1}$ & $K\times L$\\ \hline
\end{tabular}
\caption{Blocks used for calculating Xampling scheme computational cost using the matrix pencil method.}
\label{Tab:03}
\end{table}
\bibliographystyle{spiebib}   

\end{document}